# Магнитный порядок и спиновые флуктуации в низкоразмерных системах

А. А. Катанин[а,б] и В. Ю. Ирхин[а]

а) Институт физики металлов УрО РАН,
ул. С. Ковалевской 18, 620219, Екатеринбург, Россия
б) Макс-Планк Институт исследований твердого тела,
Штутгарт, Германия

Проанализирована современная теоретическая и экспериментальная ситуация в физике непроводящих низкоразмерных систем, для которых характерны низкие значения температуры магнитного перехода $T_M$ и существует развитый ближний порядок в широком температурном интервале выше $T_M$. Показана недостаточность как стандартной, так и самосогласованной спин-волновой теории для количественного описания экспериментальных данных этих систем. Рассмотрены теоретико-полевые подходы, позволяющие учесть вклад в термодинамические свойства ферро- и антиферромагнетиков от спин-флуктуационных возбуждений, пренебрегаемых в спин-волновых теориях.

# Содержание.

# 1. Введение

Исследование низкоразмерного магнетизма – важная задача современной физики твердого тела. Экспериментальный интерес к этой проблеме связан с необычными магнитными свойствами слоистых перовскитов, в том числе $Rb_2MnF_4$, $K_2NiF_4$ [1, 2], $K_2MnF_4$ [2, 3] (анизотропия «легкая ось»), $K_2CuF_4$, $NiCl_2$ $BaNi_2(PO_4)_2$ [4] (анизотропия «легкая плоскость»), органических соединений [5,6], ферромагнитных пленок, мультислоев и поверхностей [7, 8]. В последнее время интерес к низкоразмерным соединениям возрос в связи с исследованиями магнитных свойств медь-кислородных плоскостей в высокотемпературных сверхпроводниках, в том числе на основе $La_2CuO_4$ [9].

Другой класс низкоразмерных магнитных систем с локальными моментами – квазиодномерные соединения, содержащие цепочки магнитных атомов с маленьким межцепочечным обменом. К ним может быть отнесено, в частности, такое хорошо экспериментально исследованное соединение как $KCuF_3$ [10], а также ряд недавно исследованых систем на основе стронция, например $Sr_2CuO_3$ ($S = 1/2$) [11,12] и цезия: $CsNiCl_3$ ($S = 1$) [13], $CsVCl_3$ ($S = 3/2$) [14]. Родственный класс соединений представляет собой недавно синтезированные системы со «спиновыми лестницами» – ограниченным числом цепочек магнитных атомов, связанных обменным взаимодействием [15].

В отличие от трехмерных систем, возможность магнитного упорядочения в низкоразмерных системах значительно ограничена из-за сильных флуктуаций магнитного параметра порядка. Как известно, магнитный порядок в чисто одно- и двумерных изотропных системах отсутствует при конечных температурах. Согласно теореме Мермина-Вагнера, двумерные изотропные магнетики обладают дальним порядком только в основном состоянии, а точные результаты для одномерных изотропных антиферромагнетиков свидетельствуют об отсутствии дальнего магнитного порядка даже при *T*=0. Реальные соединения обладают конечной величиной температуры магнитного перехода $T_M \ll |J|$ ($J$ – величина обменного взаимодействия в цепочках или в плоскости), обусловленной слабым межцепочечным (межплоскостным) обменом и/или анизотропией. Малость температуры перехода приводит к ряду специфических особенностей этих систем. В частности, выше точки магнитного перехода ближний магнитный порядок полностью не разрушается (в двумерной ситуации он сохраняется до $T \sim |J|$), так что существует широкая область выше $T_M$ с сильным ближним порядком [3,9].

Существенный прогресс в понимании свойств основного состояния и термодинамики одно- и двумерных систем был достигнут благодаря численным методам (квантовый метод Монте-Карло и метод ренормгруппы). В то же время такие методы не заменяют аналитических подходов, позволяющих описать термодинамические свойства слоистых систем в широком интервале температур и полезных как для теоретического понимания физических свойств этих систем, не очевидных из результатов численных расчетов, так и для практических целей описания реальных соединений.

Стандартная теория спиновых волн [16-18] применима к низкоразмерным магнетикам лишь при низких температурах $T \ll T_M$. Эта теория пренебрегает взаимодействием спиновых волн, что приводит, в частности к резкому завышению температур фазового перехода низкоразмерных соединений. Проблема магнон-магнонного взаимодействия в ферромагнетиках впервые детально исследовалась в классических работах Дайсона [17], построившего последовательную теорию термодинамических свойств при низких температурах. Позже эти результаты были воспроизведены Малеевым с помощью нелинейного бозонного представления спиновых операторов [18]. В этом формализме проблема взаимодействия спиновых волн сводится к динамическому взаимодействию магнонов. Формализм Дайсона-Малеева был применен к проблеме взаимодействия спиновых волн в трехмерных [19] и двумерных [20] антиферромагнетиках; особое внимание в этих работах уделялось вычислению спин-волнового затухания, которое оказалось малым в широкой области импульсного пространства при достаточно низких температурах. Неаналитические поправки к спектру спиновых волн и теплоемкости низкоразмерных систем, возникающие за счет динамического взаимодействия магнонов, были исследованы в работах [21].

При температурах, не малых по сравнению с температурой магнитного перехода, существенную роль начинает играть кинематическое взаимодействие спиновых волн, возникающее вследствие ограничения числа бозонов на узле. Бозон-фермионное представление, позволяющее в явном виде учесть кинематическое взаимодействие спиновых волн, было предложено Барьяхтаром, Криворучко и Яблонским [22,23]. Введение вспомогательных фермионов в этом представлении позволяет избежать дополнительного условия для числа бозонов на узле. При не слишком низких температурах, однако, спин-волновая картина возбуждений становится полностью неадекватной и для правильного описания термодинамики необходим учет неспинволновых возбуждений. В некоторой степени эта ситуация аналогична теории зонного магнетизма, где теория Стонера (среднего поля) неспособна адекватно описать термодинамические свойства, что стимулировало развитие спин-флуктуационных теорий [24]. Последние оказались особенно успешными в случае слабых зонных магнетиков, аналогичных, в некоторой степени, низкоразмерным магнитным системам с малыми значениями точки перехода. В то время как вклад неспинволновых возбуждений в термодинамические свойства локализованных магнетиков обсуждался много лет назад в рамках феноменологической теории [25,26], соответствующая микроскопический подход начал развиваться лишь в последнее время в рамках так называемого $1/N$ разложения [27], где $N$ − число спиновых компонент (*N*=3 для модели Гейзенберга). Это разложение оказалось удивительно успешным при описании термодинамических свойств двумерных [27] и квазидвумерных [28] магнетиков.

В одномерных антиферромагнетиках картина спектра возбуждений сильно зависит от спина S. Начиная с работ Бете, построившего точную волновую функцию («анзатц Бете») для одномерной антиферромагнитной цепочки, известно, что эти системы не обладают дальним магнитным порядком даже в основном состоянии. Современные теоретические подходы к одномерным системам основаны на идее Халдейна [29,30], выполнившего преобразование проблемы цепочки к нелинейной сигма-модели. Согласно результатам Халдейна, случаи целого и полуцелого спина качественно различны. Для полуцелого спина появляется так называемый топологический член в эффективном действии, приводящий к необычному магнитному поведению таких цепочек.

Для одной цепочки с $S = 1/2$ (та же самая ситуация имеет место при любом полуцелом значении спина), основное состояние обладает «квазидальним порядком», когда спиновые корреляции на больших расстояниях спадают по степенному, а не экспонециальному закону. Спектр возбуждений при этом является бесщелевым, хотя намагниченность равна нулю (что напоминает двумерную классическую *XY* модель ниже точки Березинского-Костерлица-Таулеса $T_{BKT}$). В то же время для целых значений спина *S* спектр возбуждений содержит так называемую халдейновскую щель порядка exp(-π*S*) и структура спектра возбуждений близка к предсказаниям спин-волновой теории.

В связи с «экзотическим» поведением цепочек с полуцелым спином они не могут быть исследованы в рамках спин-волновой теории и их рассмотрение требует принципиально новых физическим подходов. Для предельно квантового случая $S = 1/2$ (который также наиболее важен с практической точки зрения) был развит метод бозонизации, использующий представление Йордана-Вигнера спиновых операторов через фермионные.

Далее выполняется переход от фермионных операторов к бозонным, описывающим коллективные (не спин-волновые) магнитные возбуждения. Этот подход оказался также успешен при исследовании спиновых лестниц [15,31,32].

Для исследования квазиодномерных систем были развиты комбинация бозонизации (и/или Бете-анзаца) с методом ренормгруппы [33-35] и межцепочечным приближением среднего поля [36]. Эти методы предсказывают конечную величину температуры магнитного перехода $T_N \propto |J'|$ при сколь угодно малой величине межцепочечного взаимодействия $J'$. В то время как первый подход не позволяет получить каких-либо количественных оценок величины $T_N$, второй пренебрегает спиновыми корреляциями на разных цепочках, что приводит к резкому завышению температур Нееля по сравнению с их экспериментальными значениями. Таким образом, теория межцепочечного среднего поля приводит к тем же трудностям при описании квазиодномерных магнетиков, что и спин-волновая теория в квазидвумерных магнетиках. Эта ситуация опять же аналогична проблемам теории Стонера при описании зонных магнетиков.

Итак, описание квазидвумерных и квазиодномерных магнетиков требует существенно новых подходов к этим системам, рассмотрение которых и является предметом настоящего обзора.

## 2. Квазидвумерные магнетики с анизотропией типа «легкая ось»

Для рассмотрения квазиодномерных и двумерных магнетиков с локализованными моментами используем модель Гейзенберга

$$H = -\frac{J}{2}\sum_{i\delta_\parallel} \mathbf{S}_i \mathbf{S}_{i+\delta_\parallel} + H_{3D} + H_{anis} \qquad (2.1)$$

$$H_{3D} = -\frac{J'}{2}\sum_{i\delta_\perp} \mathbf{S}_i \mathbf{S}_{i+\delta_\perp},$$

$$H_{\text{anis}} = -\frac{J\eta}{2}\sum_{i\delta_\parallel} S_i^z S_{i+\delta_\parallel}^z - |J|\zeta \sum_i (S_i^z)^2, \qquad (2.2)$$

где $J > 0$ для ферромагнетика, $J < 0$ для антиферромагнетика – обменный интеграл в плоскости, $H_{3D}$ соответствует гамильтониану межцепочного (межслоевого) взаимодействия, $J' = 2\alpha J$ является параметром обмена между цепочками (слоями), для определенности ниже рассматривается случай $\alpha > 0$, $\delta_\parallel$ и $\delta_\perp$ обозначают ближайших соседей в пределах цепочки (слоя) и для различных цепочек (слоев). $H_{\text{anis}}$ – анизотропная часть взаимодействия, возникающая в результате влияния кристаллического поля окружающих ионов; $\eta, \zeta > 0$ — соответственно параметры обменной и одноионной анизотропии соответственно.

## 2.1. Нелинейные бозонные представления в теории квазидвумерных ферро- и антиферромагнетиков

При достаточно низких температурах $T \ll T_M$ элементарными возбуждениями в магнетиках являются спиновые волны. Для описания этих возбуждений удобно перейти от спиновых операторов к бозонным. В настоящее время используются различные представления такого вида, в частности представление Дайсона-Малеева [17,18,23]

$$S_i^+ = \sqrt{2S}\, b_i,\ S_i^z = S - b_i^\dagger b_i, \qquad (2.3)$$

$$S_i^- = \sqrt{2S}\,(b_i^\dagger - \frac{1}{2S} b_i^\dagger b_i^\dagger b_i),$$

($b_i^\dagger, b_i$ – магнонные бозе-операторы), которое удобно для описания магнитоупорядоченной фазы. Бозонные операторы в этом представлении должны удовлетворять условию на числа заполнения на узле $N_{bi} < 2S$, что приводит к так называемому кинематическому взаимодействию спиновых волн. Чтобы обойти эту трудность, Барьяхтар, Криворучко и Яблонский ввели представление [22,23]

$$S_i^+ = \sqrt{2S}\,b_i,\ S_i^z = S - b_i^\dagger b_i - (2S+1)c_i^\dagger c_i, \qquad (2.4)$$

$$S_i^- = \sqrt{2S}(b_i^\dagger - \frac{1}{2S}b_i^\dagger b_i^\dagger b_i) - \frac{2(2S+1)}{\sqrt{2S}}b_i^\dagger c_i^\dagger c_i,$$

содержащее помимо бозонных операторов вспомогательные псевдофермионные операторы $c_i^\dagger, c_i$, учитывающие кинематическое взаимодействие спиновых волн. В случае антиферромагнетика с двумя подрешетками используется разбиение исходной решетки на две подрешетки, в каждой из которых используется представление (2.4) и сопряженное ему. При низких температурах соответствующая энергия псевдофермионов порядка |*J*|, так что их вклад в термодинамические величины экспонециально мал и им можно пренебречь. В то же время, кинематическое взаимодействие спиновых волн становится существенным при *T*~|*J*|.

Другое полезное представление спиновых операторов – представление швингеровских бозонов [37-39]

$$\mathbf{S}_i = \frac{1}{2}\sum_{\sigma\sigma'} s_{i\sigma}^\dagger \boldsymbol{\sigma}_{\sigma\sigma'} s_{i\sigma'}, \qquad (2.5)$$

где $\boldsymbol{\sigma}$ – матрицы Паули, $\sigma, \sigma' = \uparrow, \downarrow$, так что

$$S_i^z = \frac{1}{2}(s_{i\uparrow}^\dagger s_{i\uparrow} - s_{i\downarrow}^\dagger s_{i\downarrow}),\ S_i^+ = s_{i\uparrow}^\dagger s_{i\downarrow},\ S_i^- = s_{i\downarrow}^\dagger s_{i\uparrow}. \qquad (2.6)$$

Условие

$$s_{i\uparrow}^\dagger s_{i\uparrow} + s_{i\downarrow}^\dagger s_{i\downarrow} = 2S \qquad (2.7)$$

ограничивает число спиновых состояний и должно выполняться на каждом узле решетки. Так как одновременное изменение фаз $s_{i\uparrow}$ и $s_{i\downarrow}$ бозонов, $s_{i\sigma} \to s_{i\sigma}\exp(i\phi_i)$ не меняет физических результатов, это представление обладает калибровочной симметрией. Этот факт может быть использован для нахождения связи представления швингеровских бозонов с известным представлением Гольштейна-Примакова [40]. Действительно, если фиксировать калибровку условием эрмитовости одного из операторов $s_{i\sigma}$, например $s_{i\uparrow}$, имеем из (2.7)

$$s_{i\uparrow} = \sqrt{2S - s_{i\downarrow}^\dagger s_{i\downarrow}}. \tag{2.8}$$

Подставляя в (2.6), получаем представление Гольштейна-Примакова. Таким образом, представления швингеровских бозонов и Гольштейна-Примакова эквивалентны. Эта эквивалентность, однако, может быть нарушена в приближенных подходах. В отличие от представления Гольштейна-Примакова (или Дайсона-Малеева), представление швингеровских бозонов может быть легко обобщено на произвольное число сортов бозонов $N \geq 2$, что приводит к модели с SU($N$)/SU($N$–1) симметрией и позволяет построение $1/N$ – разложения [38]. В то же время нет никакого естественного способа введения фермиевских-операторов в это представление, чтобы учесть кинематическое взаимодействие. Как и в представлении Дайсона-Малеева, для антиферромагнетика необходимо перейти к локальной системе координат заменой [39]

$$s_{i\uparrow} \to -s_{i\downarrow},\ s_{i\downarrow} \to s_{i\uparrow}$$

в одной из двух подрешеток.

## 2.2. ССВТ квазидвумерных магнетиков

Взаимодействие магнонов в наинизшем (борновском) приближении рассматриваются в так называемой самосогласованной спин-волновой теории (ССВТ). Впервые эта теория была применена много лет назад к трехмерной модели Гейзенберга [41]; те же самые результаты были получены позднее в рамках вариационного подхода для изотропной [42] и анизотропной [43] модели Гейзенберга. Близкие идеи использовались недавно для описания двумерных магнетиков в теории "среднего поля" для бозонных операторов [38,39,44], основанной на представлении операторов спина через швингеровские бозоны, и "модифицированной спин-волновой теории" [45], основанной на представлении Дайсона-Малеева. Результаты этих теорий находятся в хорошем согласии с ренормгрупповыми вычислениями [46,47] и экспериментальными данными для спектра возбуждений низкоразмерных систем [4]. ССВТ также применялась к квазидвумерным [48-51], фрустрированным двумерным [52-56] и трехмерным [53] антиферромагнетикам.

Для вывода уравнений ССВТ используем представление Дайсона-Малеева (2.3). После подстановки в гамильтониан представления спиновых операторов через бозонные, возникают члены второй и четвертой степени по бозонным операторам. В то время как квадратичные вклады описывают распространение свободных спиновых волн, вторые соответствуют их взаимодействию. Учитывая взаимодействие спиновых волн в наинизшем приближении, т.е. расцепляя четверные формы бозонных операторов по теореме Вика, получаем квадратичный гамильтониан ССВТ

$$H = \sum_{i\delta} J_\delta \gamma_\delta \left( b_i^\dagger b_i - b_{i+\delta}^\dagger b_i \right) - \mu \sum_i b_i^\dagger b_i, \qquad (2.9)$$

где

$$\gamma_{\delta_\perp} = \gamma = \overline{S} + \langle b_i^\dagger b_{i+\delta_\perp} \rangle, \quad \gamma_{\delta_\parallel} = \gamma' = \overline{S} + \langle b_i^\dagger b_{i+\delta_\parallel} \rangle \qquad (2.10)$$

– параметры ближнего порядка, удовлетворяющие уравнениям

$$\gamma = \overline{S} + \sum_{\mathbf{k}} N_{\mathbf{k}} \cos k_x, \, \gamma' = \overline{S} + \sum_{\mathbf{k}} N_{\mathbf{k}} \cos k_z, \qquad (2.11)$$

Для распространения теории в разупорядоченную фазу в (2.9) введен химический потенциал бозонов $\mu$, дающий возможность удовлетворить условию ограниченности общего числа бозонов при $T > T_C$, где $\overline{S} = 0$ [23,30,45]. При $T < T_C$ имеем $\mu = 0$, так что число бозонов не ограничено. Вычисление спиновых корреляционных функций показывает [45], что химический потенциал непосредственно определяет корреляционную длину $\xi_\delta$ в направлении $\delta$ согласно соотношению

$$\xi_\delta^{-1} = \sqrt{-\mu/|J_\delta \gamma_\delta|}. \qquad (2.12)$$

Величины $\gamma$ и $\gamma'$ определяют спиновые корреляционные функции на соседних узлах,

$$|\langle \mathbf{S}_i \mathbf{S}_{i+\delta} \rangle| = \gamma_\delta^2. \qquad (2.13)$$

Намагниченность ферромагнетика определяется полным числом бозонов:

$$\overline{S} = S - \sum_{\mathbf{k}} N_{\mathbf{k}}, \qquad (2.14)$$

где $N_{\mathbf{k}} = N(E_{\mathbf{k}})$ – функция Бозе, причем спектр спиновых волн имеет вид

$$E_{\mathbf{k}}^{\text{SSWT}} = \Gamma_0 - \Gamma_{\mathbf{k}} + \Delta - \mu, \qquad (2.15)$$

$$\Gamma_{\mathbf{k}} = 2S[\gamma |J|(\cos k_x + \cos k_y) + \gamma' |J'|\cos k_z],$$

$$\Delta = |J|\left[(2S-1)\zeta + 4\eta S^2/\gamma\right](\overline{S}/S)^2.$$

Хотя ССВТ может быть обоснована лишь при температурах $T \ll T_M$ (при которых существует развитый дальний порядок и взаимодействие спиновых волн мало), представляет интерес экстраполяция результатов ССВТ на более высокие температуры $T \sim T_M$, что позволяет сравнить результаты ССВТ с результатами более сложных теорий, рассматриваемых в следующем разделе.

Для антиферромагнетика уравнения ССВТ имеют вид [50,51]

$$\gamma = \overline{S} + \sum_{\mathbf{k}} \frac{\Gamma_{\mathbf{k}}}{2E_{\mathbf{k}}} \cos k_x \coth \frac{E_{\mathbf{k}}}{2T}, \qquad (2.16)$$

$$\gamma' = \overline{S} + \sum_{\mathbf{k}} \frac{\Gamma_{\mathbf{k}}}{2E_{\mathbf{k}}} \cos k_z \coth \frac{E_{\mathbf{k}}}{2T},$$

$$\overline{S} = S + 1/2 - \sum_{\mathbf{k}} \frac{\Gamma_0 + \Delta - \mu}{2E_{\mathbf{k}}} \coth \frac{E_{\mathbf{k}}}{2T},$$

где

$$\gamma = \overline{S} + \langle a_i b_{i+\delta_\perp} \rangle, \quad \gamma' = \overline{S} + \langle a_i b_{i+\delta_\parallel} \rangle \qquad (2.17)$$

и энергия спиновых волн равна

$$E_{\mathbf{k}}^{\text{SSWT}} = [(\Gamma_0 + \Delta - \mu)^2 - \Gamma_{\mathbf{k}}^2]^{1/2}. \qquad (2.18)$$

Как и для ферромагнетиков, химический потенциал бозонов $\mu$, отличный от нуля выше температуры магнитного перехода, определяет корреляционную длину согласно соотношению (2.12).

В основном состоянии ферромагнетика $\bar{S}_0 = S$ и $\gamma_0 = \gamma(T=0) = 1$, подрешеточная намагниченность и параметр ближнего порядка двумерного антиферромагнетика отличается от этих значений из-за квантовых нулевых колебаний спинов:

$$\bar{S}_0 = S - \frac{1}{2}\sum_{\mathbf{k}}\left[\frac{1}{\sqrt{1-\phi_{\mathbf{k}}^2}} - 1\right] \simeq S - 0.1966, \qquad (2.19)$$

$$\gamma_0 = 1 + \frac{1}{2S}\sum_{\mathbf{k}}\left[1 - \sqrt{1-\phi_{\mathbf{k}}^2}\right] \approx 1 + \frac{0.0790}{S}, \qquad (2.20)$$

где $\phi_{\mathbf{k}} = \frac{1}{2}(\cos k_x + \cos k_y)$. При этом при S=1/2 подрешеточная намагниченность составляет 40% от ее величины в ферромагнитном случае и совпадает с ее значением в спин-волновой теории [16], величина перенормировки параметра обмена в плоскости равна 15%. Как и в стандартной теории спиновых волн, в отсутствие анизотропии ($\delta = 0$) спектр спиновых волн в упорядоченной фазе является бесщелевым и при малых $k$ имеет вид $E_k = Dk^2$ в ФМ случае и $E_k = ck$ в АФМ случае, где $D$ – константа жесткости спиновых волн, $c$ – скорость спиновых волн. В ССВТ эти параметры выражаются через параметры (2.19), (2.20) согласно соотношениям

$$D = JS, \quad c = \sqrt{8}\,|J|\gamma S. \qquad (2.21)$$

Выражение для спиновой жесткости ферро- и антиферромагнетика, определенной из анализа поперечной восприимчивости, имеет вид

$$\rho_s = JS^2 \,(\text{ФМ}), \quad \rho_s = |J|\gamma S \bar{S}_0 \,(\text{АФМ}) \qquad (2.22)$$

Перенормированные (наблюдаемые) параметры межплоскостного обмена и анизотропии, определенные из спектра возбуждений равны

$$f_r = \frac{\Delta}{\gamma\,|J|\,S} = \frac{1}{\gamma S}\left[(2S-1)\zeta + 4\eta S/\gamma\right](\bar{S}/S)^2, \qquad (2.23)$$

$$\alpha_r = \frac{2\gamma'}{\gamma} = \alpha \bar{S}/S. \qquad (2.24)$$

Отметим, что в отличие от параметра внутриплоскостного обмена, перенормированные параметры $\alpha, \eta, \zeta$ пропорциональны намагниченности, и, таким образом, обладают сильной температурной зависимостью.

При конечных температурах в отсутствии межплоскостного обмена и анизотропии ($J^{'} = 0$, $\delta = 0$) дальний порядок отсутствует в соответствии с теоремой Мермина-Вагнера, так что $\overline{S} = 0$, $\mu < 0$ (решения с $\overline{S} \neq 0$, $\mu = 0$ отсутствуют вследствие расходимости интегралов в уравнениях (2.11) и (2.16) при $T > 0$ и $\mu = 0$). Величина химического потенциала бозонов µ определяется для ферромагнетика уравнением (2.14), для антиферромагнетика – последним из уравнений (2.16) с $\overline{S} = 0$. При низких температурах $T \ll |J|S^2$ абсолютная величина химического потенциала экспоненциально мала, так что корреляционная длина $\xi = \sqrt{-|J\gamma|/\mu}$ экспоненциально велика (так называемый перенормированный классический режим),

$$\xi = C_\xi^{\text{F}} \sqrt{J/T} \exp(2\pi\rho_s/T) \quad \text{(ФМ)}, \tag{2.25}$$

$$\xi = C_\xi^{\text{AF}} (J/T) \exp(2\pi\rho_s/T) \, \text{(АФМ)}, \tag{2.26}$$

где $C_\xi^{\text{F,AF}}$ – зависящие от спина константы. Результаты (2.25), (2.26) согласуются с результатами однопетлевого ренормгруппового (РГ) подхода [46,47]. Двухпетлевой РГ анализ изменяет только предэкспонециальный множитель: в АФМ случае он становится температурно-независимой постоянной [46], в то время как в ФМ случая пропорционален $(T/J)^{1/2}$ (см. [47]).

В присутствии межплоскостного обмена при не слишком высоких температурах $T < T_M$ (температура магнитного упорядочения $T_M$ будет рассчитана ниже) появляется дальний магнитный порядок, при этом уравнения (2.11) и (2.16) имеют решения с $\overline{S} > 0$.

При низких температурах ($T \ll |J^{'}|S$) и произвольном $J^{'}/J$ поправки к намагниченности основного состояния ферромагнетика пропорциональны $T^{3/2}$, в то время как параметры ближнего порядка имеют более слабую $T^{5/2}$ – зависимость, для антиферромагнетика соответствующие зависимости – $T^2$ и $T^4$ [51]. При $T > T_M$ снова имеем $\bar{S} = 0$ и $\mu < 0$, так же как в двумерном случае при конечных $T$.

Для численного исследования температурной зависимости намагниченности и параметров ближнего порядка при не слишком малых значениях межслоевого обмена удобно использовать приближение эффективного параметра ближнего порядка, производя замену [51]

$$\sum_\delta J_{i,i+\delta}\gamma_\delta(b_i^\dagger b_i - b_i^\dagger b_{i+\delta}) \to \gamma_{\text{ef}}\sum_\delta J_{i,i+\delta}(b_i^\dagger b_i - b_i^\dagger b_{i+\delta}). \qquad (2.27)$$

Температурная зависимость намагниченности и параметра ближнего порядка ферромагнетика для различных $J'/J$ показаны на рис. 1-3. При малых $T - T_M$ имеем $-\mu \propto (T - T_M)^2$ (см. рис. 3 для ферромагнитного случая, та же самая ситуация имеет место в АФМ случае), так что согласно критический индекс для корреляционной длины $\nu = 1$. Так как намагниченность изменяется линейно около $T_M$, критический индекс намагниченности $\beta = 1$. Влияние поправок более высокого порядка по $1/S$ на значение критических индексов обсуждается ниже. В классическом пределе $S \to \infty$ уравнения ССВТ упрощаются и при $T < T_M$ ($\mu = 0$) усредненный (по ближайшим соседям) параметр ближнего порядка

$$\gamma_{\text{ef}}(T) = (4J\gamma + 2J'\gamma')/J_0 \qquad (2.28)$$

(но не намагниченность!) удовлетворяет стандартному уравнению среднего поля

$$\gamma_{\text{ef}}/S = B_\infty\left(J_0\gamma_{\text{ef}} S/T\right), \qquad (2.29)$$

где $B_\infty(x) = \coth x - 1/x$ – функция Ланжевена (функция Бриллюэна в классическом пределе). Температура $T^*$, при которой $\gamma_{\text{ef}}(T^*) = 0$, оказывается выше чем температура магнитного фазового перехода $T_M$, так что $\gamma_{ef}(T_M) > 0$, а поведение $\gamma_{ef}$ при $T > T_M$ более сложно, чем (2.29).

При малых значениях межплоскостного обмена $J'/J \ll 1$ и анизотропии $\eta, \zeta \ll 1$ возможно получение аналитических результатов для температурной зависимости намагниченности в широком диапазоне температур [48,51]. При этом ССВТ приводит к различным результатам для намагниченности в так называемом «квантовом» и «классическом» температурных режимах. Оказывается, что эти режимы не связаны однозначно со случаем квантовых ($S \sim 1$) и классических ($S \gg 1$) спинов (хотя классический режим реализуется лишь при $S \gg 1$), поскольку реальные критерии содержат также температуру (см. ниже).

В квантовом режиме, который имеет место при не слишком низких температурах

$$J'S \ll T \ll JS \quad (\text{ФМ}),$$
$$(JJ')^{1/2} S \ll T \ll |J|S \quad (\text{АФМ}) \qquad (2.30)$$

(подрешеточная) намагниченность равна

$$\overline{S} = \begin{cases} S - \dfrac{T}{4\pi JS} \ln \dfrac{T}{J'\gamma'S} & (\text{ФМ}), \\ \overline{S}_0 - \dfrac{T}{4\pi |J|\gamma S} \ln \dfrac{T^2}{8JJ'\gamma\gamma'S^2} & (\text{АФМ}) \end{cases} \qquad (2.31)$$

Параметры ближнего порядка определяются соотношениями $\gamma \simeq \gamma_0$ и

$$\gamma' = \begin{cases} S - \dfrac{T}{4\pi JS}\left(\ln \dfrac{T}{J'\gamma'S} - 1\right) & (\text{ФМ}), \\ \overline{S}_0 - \dfrac{T}{4\pi |J|\gamma S}\left(\ln \dfrac{T^2}{8JJ'\gamma\gamma'S^2} - 1\right) & (\text{АФМ}), \end{cases} \qquad (2.32)$$

так что $\gamma'_0 = \overline{S}_0$. Отметим, что в квантовом режиме (2.30) интегралы по квазиимпульсам в уравнениях ССВТ определяются вкладом квазиимпульсов $q < q_0$, где

$$q_0 = \begin{cases} (T/JS)^{1/2} & (\text{ФМ}) \\ T/c & (\text{АФМ}) \end{cases}, \qquad (2.33)$$

а не всей зоной Бриллюэна. Для критических температур в режиме (2.30) получаем результаты

$$T_C = \frac{4\pi J S^2}{\ln(T/J'\gamma_c'S)}, \qquad (2.34)$$

$$T_N = \frac{4\pi |J| \gamma_c \overline{S}_0}{\ln(T^2/8JJ'\gamma_c\gamma_c'S^2)},$$

где $\gamma_c \simeq \gamma_0$ и $\gamma_c'$ – перенормированные обменные параметры в $T_M = T_C(T_N)$; значение $\gamma_c'$, определенное из (2.32), есть

$$\gamma_c' = (T_M/4\pi |J| \gamma_c S^2) J'. \qquad (2.35)$$

Перенормировка межплоскостного обмена в (2.34) приводит к существенному понижению температуры Кюри (Нееля) по сравнению с ее значением в спин-волновой теории, поскольку $\gamma_c\gamma_c'/JJ' = T_M/4\pi JS^2 \ll 1$.

В случае больших $S$ (классический предел) получаем для ферро- и антиферромагнетиков при $T \gg |J| S$

$$\overline{S} = S - \frac{T}{4\pi |J| S} \ln \frac{q_0^2 J}{J'\gamma'}, \qquad (2.36)$$

$$\gamma' = S - \frac{T}{4\pi |J| S} \left( \ln \frac{q_0^2 J}{J'\gamma'} - 1 \right). \qquad (2.37)$$

В отличие от квантового случая, результаты для намагниченности в этом пределе неуниверсальны, так как зависят от типа решетки через параметр обрезания $q_0^2$ (для квадратной решетки $q_0^2 = 32$). Соответствующее выражение для критической температуры классического магнетика с $1 \ll \ln(q_0^2 J/J') \ll 2\pi S$ имеет вид

$$T_M = \frac{4\pi |J| S^2}{\ln(q_0^2 J/J'\gamma_c')}, \qquad (2.38)$$

где $\gamma_c' = T_M/4\pi |J| S$. При этом критическая температура одинакова для классических ферро- и антиферромагнетиков. С логарифмической точностью в этом случае воспроизводятся результаты спин-волновой теории, где $\gamma_c'/S \to 1$.

Аналогично ситуации малого межплоскостного обмена, выделение логарифмических особенностей при малой анизотропии приводит к результатам

$$\overline{S} = S - \frac{T}{4\pi JS}\ln\frac{T}{\Delta}, \qquad (\text{ФМ}), \qquad (2.39)$$

$$\overline{S} = \overline{S}_0 - \frac{T}{4\pi |J|\gamma S}\ln\frac{T^2}{8\gamma JS\Delta}, \quad (\text{АФМ}).$$

В отличие от квазидвумерного случая, здесь возникает нефизический результат $\Delta(T_M) = 0$ из-за пропорциональности щели магнитного спектра $(\overline{S}/S)^2$ (на самом деле конечная величина щели при $T = T_M$ связана с топологическими возбуждениями - «доменными стенками», которые не учитываются спин-волновой теорией). Таким образом, спин-волновая теория неспособна описать зависимость щели $\Delta(T)$ вблизи $T_M$. Обозначая формально $\Delta_c = \Delta(T_M)$, имеем для критической температуры при $2\pi S \ll \ln(1/\Delta)$

$$T_C = \frac{4\pi JS^2}{\ln(T/\Delta_c)}, \qquad (2.40)$$

$$T_N = \frac{4\pi |J| S\overline{S}_0 \gamma_c}{\ln(T^2/8J\gamma_c S\Delta_c)}.$$

В пределе больших $S$ находим как для ферро-, так и для антиферромагнетиков

$$\overline{S} = S - \frac{T}{4\pi |J| S}\ln\frac{|J| Sq_0^2}{\Delta}. \qquad (2.41)$$

Намагниченность (2.41) исчезает при температуре

$$T_M = \frac{4\pi |J| S^2}{\ln(|J| Sq_0^2/\Delta_c)}. \qquad (2.42)$$

соответствующей критической температуре классического магнетика с анизотропией типа «легкая ось» ($1 \ll \ln(|J| q_0^2/\Delta) \ll 2\pi S$).

Результаты (2.34) и (2.38) могут быть сопоставлены с результатом приближения Тябликова для температуры магнитного перехода слоистых соединений [57, 28],

$$T_M \simeq \frac{4\pi |J| S^2}{\ln(|J| q_0^2 / J^{'})} \qquad (2.43)$$

с $q_0^2 = 32$. Результат (2.43) численно меньше, чем значение ССВТ (2.34), а потому лучше описывает экспериментальные данные (см. раздел 2.6). С другой стороны, (2.43) совпадает с результатом для сферической модели (который является адекватным только в классическом пределе $S \to \infty$ [64]) и с результатом приближения спиновых волн в классическом режиме. Это показывает, что вблизи критической температуры приближение Тябликова не учитывает квантовые флуктуации, которые важны при малых значениях $S$. Таким образом, приближение Тябликова может быть частично удовлетворительно с практической, но не с теоретической точки зрения.

Хотя с логарифмической точностью все обсуждавшиеся подходы приводят к одному и тому же значению температуры Нееля, эта точность недостаточна для количественного описания экспериментальных данных, критическое поведение описывается спин-волновыми теориями также неправильно. Формально ССВТ соответствует пределу $N \to \infty$ в SU(*N*)/SU(*N*–1) обобщении модели Гейзенберга [38]. Чтобы улучшить описание критической области и вычисление температур Кюри (Нееля), необходимо рассмотреть флуктуационные поправки к результатам теории спиновых волн более аккуратно, чем в ССВТ. Вычисление поправок первого порядка по 1/*N* в SU(*N*)/SU(*N*–1) модели может позволить описать область низких и промежуточных температур $T \lesssim T_M$, но неспособно правильно описать критическое поведение [58]. Проблемы этого подхода в критической области связаны с тем, что в указанном обобщении модели Гейзенберга возбуждения неспинволнового характера представляются как связанные состояния спиновых волн [59] и их рассмотрение весьма затруднительно в рамках 1/*N* разложения. В связи с этим, необходимо развитие подходов, позволяющих описать как область промежуточных температур, так и критическую область. Такие подходы рассматриваются ниже.

## 2.3. Теоретико-полевое описание квазидвумерных магнетиков с локализованными моментами

Для правильного описания термодинамических свойств в широком интервале температур необходимо суммирование ведущих вкладов в термодинамические величины во всех порядках теории возмущений по магнон-магнонному взаимодействию. При этом кинематическое взаимодействие спиновых волн важно в широкой температурной области только для систем, где $T_M$ не мало по сравнению с $|J|S^2$ (например, для трехмерных систем). Для слоистых систем кинематическое взаимодействие менее важно в силу малости температуры магнитного перехода $T_M \ll |J|S^2$ (фактически, это взаимодействие играет роль только в узкой критической области около $T_M$).

В то же время принципиально важным является правильный учет динамического магнон-магнонного взаимодействия. Для суммирования диаграмм, описывающих влияние этого взаимодействия за пределами низшего порядка теории возмущений возможно применение ренормгруппового (РГ) анализа. Этот подход был успешно применен для описания классических и квантовых изотропных магнетиков в пространствах размерности $d=2$ [46,60] и $d=2+\varepsilon$ [61,62]. В указанных случаях картина спектра возбуждений слабо отличается от спин-волновой, так при $d=2+\varepsilon$ поправки к спектру спиновых волн $\delta E_\mathbf{q} \sim |J|\varepsilon \ln q$. При этом температура магнитного перехода $T_M/|J|S^2 \sim \varepsilon$ и может быть применена стандартная техника $\varepsilon$-разложения [61,62]. Соответствующие результаты РГ анализа совпадают с результатами $1/N$ разложения в SU($N$)/SU($N-1$) обобщении модели Гейзенберга [58].

В случае квазидвумерных магнетиков со слабым межплоскостным обменом и/или слабой анизотропией типа «легкая ось» температура магнитного перехода также мала в сравнении с $|J|S^2$, однако спектр возбуждений может существенно отличаться от спин-волнового. В частности, можно выделить три температурных режима [63]. При низких температурах $T \ll T_M$ применима спин-волновая теория. При промежуточных температурах $T \sim T_M$ (вне критической области) взаимодействие спиновых волн становится существенным, но спиновые флуктуации носят двумерный изотропный характер (по этой причине этот режим далее именуется «двумерный гейзенберговский режим»); для описания магнитных свойств в этом режиме может быть применен метод РГ. Наконец, в узкой критической области вблизи $T_M$ происходит переход от вышеупомянутого двумерного гейзенберговского режима к трехмерному гейзенберговскому *критическому* режиму, в котором существенны флуктуации неспинволнового характера. В присутствии анизотропии флуктуации в критическом режиме обусловлены наличием топологических возбуждений типа доменных стенок, т.е. являются «двумерными изинговскими». В обоих случаях в достаточно узкой критической области картина спиновых волн становится полностью неадекватной. Таким образом, эта область должна рассматриваться с учетом существенно неспинволновых возбуждений.

Для рассмотрения этого типа возбуждений удобно использовать вместо исходной модели Гейзенберга так называемую $O(N)/O(N-1)$ модель с большим числом компонент спина $N$, позволяющую ввести в теорию формально малый параметр [27]. При $N=\infty$ указанная модель эквивалентна сферической модели [64], однако при конечных значениях $N$ она правильно учитывает поправки связанные со спин-спиновым взаимодействием, поскольку не основана на спин-волновой картине спектра. Это обстоятельство приводит к важным преимуществам при температурах, сравнимых с температурой фазового перехода, но ведет к некоторым трудностям при описании низких и промежуточных температур, где возбуждения имеют чисто спин-волновой характер. Таким образом, подходы РГ и $1/N$ разложения в $O(N)/O(N-1)$ модели имеют преимущества в различных температурных областях и взаимно дополняют друг друга. В то время как первый метод хорошо описывает двумерный режим, реализующийся при промежуточных температурах, $1/N$ разложение удовлетворительно описывает критическую область.

Применения РГ подхода и $1/N$ разложения предполагает переформулировку исходной проблемы вычисления термодинамических свойств модели с гамильтонианом (2.1) на языке континуального интеграла [65,66]. В этом формализме модель характеризуется так называемым производящим функционалом $Z[h]$, являющемся статистической суммой системы во внешнем неоднородном магнитном поле $h$, так что логарифмические производные $Z[h]$ по полю определяют намагниченность и корреляционные функции. Для вывода производящего функционала используется представление когерентных состояний [65,66]

$$|\mathbf{n}_i\rangle = \exp(-i\varphi_i S_i^z)\exp(-i\theta_i S_i^y)|0\rangle \qquad (2.44)$$

параметризуемых векторами $\mathbf{n}_i$ единичной длины с полярными координатами $(\theta_i, \varphi_i)$, определенных для каждого узла решетки $i$, $|0\rangle$ - собственное состояние оператора $S_i^z$ с максимальной проекцией спина: $S_i^z|0\rangle = S|0\rangle$. Преимущество состояний (2.44) состоит в том, что среднее значение операторов спина в них имеет простой вид:

$$\langle \mathbf{n}_i | S_i^m | \mathbf{n}_i \rangle = S\, n_i^m \qquad (2.45)$$

т.е. когерентные состояния являются «квазиклассическими» спиновыми состояниями. Можно показать что для когерентных состояний (2.44) производящий функционал может быть записан в виде

$$Z = \int D\mathbf{n}\exp\left\{\int_0^{1/T} d\tau\left[\mathbf{A}(\mathbf{n}_i)\frac{\partial \mathbf{n}_i}{\partial \tau} - \langle \mathbf{n}|H|\mathbf{n}\rangle\right]\right\} \qquad (2.46)$$

где первый член в показателе экспоненты учитывает динамику спинов, связанную с их квантовым характером, а второй член описывает взаимодействие спинов; интегрирование в (2.46) производится по угловым переменным вектора $\mathbf{n}_i$ на каждом узле и для каждого момента мнимого времени τ. $\mathbf{A}(\mathbf{n})$ – векторный потенциал единичного магнитного монополя, удовлетворяющий соотношению $\nabla \times \mathbf{A}(\mathbf{n}) \cdot \mathbf{n} = 1$. Вклад в действие с производной по времени соответствует фазе Берри [29].

Среднее по когерентным состояниям гамильтониана (2.1) может быть легко вычислено с учетом соотношений (2.45) и приводит к выражению для производящего функционала модели (2.1) в виде

$$Z[h] = \int D\mathbf{n} D\lambda \exp\left\{\frac{JS^2}{2} \int_0^{1/T} d\tau \sum_{i,\delta} \left[\frac{2\mathrm{i}}{JS}\mathbf{A}(\mathbf{n}_i)\frac{\partial \mathbf{n}_i}{\partial \tau} + \mathbf{n}_i \mathbf{n}_{i+\delta_\parallel} \right.\right.$$
$$\left.\left. + \frac{\alpha}{2}\mathbf{n}_i \mathbf{n}_{i+\delta_\perp} + \eta n_i^z n_{i+\delta_\parallel}^z + \mathrm{sgn}(J)\widetilde{\zeta}(n_i^z)^2 + h n_i^z + \mathrm{i}\lambda_i(\mathbf{n}_i^2 - 1)\right]\right\}, \qquad (2.47)$$

где $\widetilde{\zeta} = 2\zeta(1 - 1/2S)$. Функционал (2.47) является обобщением хорошо известного производящего функционала двумерной изотропной модели Гейзенберга [66] при введении межплоскостного обмена (член пропорциональный α) и анизотропии (члены пропорциональные $\eta$ и $\widetilde{\zeta}$). Последнее слагаемое в показателе экспоненты возникает вследствие ограничения $\mathbf{n}_i^2 = 1$. Функционал (2.47) содержит две переменные с размерностью длины:

$$\xi_{J'} = a / \max(\alpha, \widetilde{\zeta}, \eta)^{1/2} \gg a \qquad (2.48)$$

и

$$L_\tau = \begin{cases} a\sqrt{JS/T} & (\text{ФМ}) \\ c/T & (\text{АФМ}) \end{cases}. \qquad (2.49)$$

На масштабе $\xi_{J'}$ характер флуктуаций изменяется с двумерных гейзенберговского типа на трехмерные гейзенберговские или двумерные изинговские флуктуации в зависимости от того, какой из параметров доминирует в знаменателе (2.48) – анизотропия или межплоскостной обмен. Величина $L_\tau$ определяет роль квантовых спиновых флуктуаций - $L_\tau \ll a$ соответствует классическому пределу, при котором динамикой спинов можно пренебречь, $L_\tau > a$ - квантовому.

Производящий функционал (2.47) может быть далее преобразован к виду, удобному для конкретных вычислений; при этом результат определяется температурным режимом, в котором производятся вычисления. В классическом режиме $T \gg JS$ имеем $L_\tau \ll a$ и динамикой поля ***n*** (т.е. его зависимостью от мнимого времени) можно пренебречь, что приводит к функционалу

$$Z_{\text{cl}}[h] = \int D\mathbf{n} D\lambda \exp\left\{\frac{\rho_s^0}{2T}\sum_i \left[\mathbf{n}_i \mathbf{n}_{i+\delta_\parallel} + \frac{\alpha}{2}\mathbf{n}_i \mathbf{n}_{i+\delta_\perp}\right.\right. \quad (2.50)$$

$$\left.\left. + \eta n_i^z n_{i+\delta_\parallel}^z + \widetilde{\zeta}(n_i^z)^2 + h n_i^z + \mathrm{i}\,\lambda(\mathbf{n}_i^2 - 1)\right]\right\}$$

с «затравочной» спиновой жесткостью $\rho_s^0 = |J|S^2$. Чтобы получить (2.50) в антиферромагнитном случае, необходимо произвести замену $\mathbf{n}_i \to -\mathbf{n}_i, \lambda_i \to -\lambda_i$ для одной из двух подрешеток. Таким образом, в классическом случае результаты для $Z$ идентичны для ферро- и антиферромагнетиков. В континуальном пределе действие (2.50) совпадает с действием для классической нелинейной сигма-модели [66]. Однако, если интересоваться термодинамикой в широком интервале температур (не только в критической области), континуальный предел не может быть использован, так как при этом вклад в термодинамические свойства дают не только длинноволновые, но и коротковолновые возбуждения.

В квантовом случае в силу условия $\xi_{J'} \gg L_\tau$ можно перейти к континуальному пределу для каждого слоя. Для ферромагнетика удобно использовать представление [65,66]

$$\mathbf{A}(\mathbf{n}) = \frac{\mathbf{z} \times \mathbf{n}}{1 + (\mathbf{zn})} \quad (2.51)$$

($\mathbf{z}$ – единичный вектор вдоль оси $z$) и ввести двухкомпонентное векторное поле $\pi = \mathbf{n} - (\mathbf{nz})\mathbf{z}$ описывающее флуктуации параметра порядка. Для квантового антиферромагнетика необходимо использовать процедуру Халдейна [29] (см. также [66]), чтобы проинтегрировать по «быстрым» компонентам поля **n**. Для этого вектор **n** представляется в виде

$\mathbf{n}_i = \mathbf{L}_i + (-1)^i \boldsymbol{\sigma}_i$ где $\mathbf{L}_i$ и $\boldsymbol{\sigma}_i$ - однородная («быстрая») и подрешеточная («медленная») компоненты вектора, $\mathbf{L}_i \boldsymbol{\sigma}_i = 0$. Для отделения «быстрых» и «медленных» компонент ниже точки перехода вместо обычно используемой корреляционной длины, равной бесконечности, используется параметр $\xi_{J'} \gg a$. С помощью указанной процедуры приходим к производящему функционалу квантовой нелинейной сигма-модели

$$Z_{\mathrm{AF}}[h] = \int D\sigma D\lambda \exp\left\{-\frac{\rho_s^0}{2}\int_0^{1/T} d\tau \int d^2\mathbf{r} \sum_{i_z} \left[\frac{1}{c_0^2}(\partial_\tau \sigma_{i_z})^2 \right.\right.$$
$$\left.\left. + (\nabla \sigma_{i_z})^2 + \frac{\alpha}{2}(\sigma_{i_z+1} - \sigma_{i_z})^2 - f(\sigma_{i_z}^z)^2 + h\sigma_{i_z}^z + i\lambda(\sigma_{i_z}^2 - 1)\right]\right\}, \qquad (2.52)$$

где $\sigma_{i_z}$ – трехкомпонентное поле единичной длины и $c_0 = \sqrt{8}JS$ – затравочная скорость спиновых волн. Модель (2.52) является обобщением квантовой нелинейной сигма-модели на случай наличия межплоскостного обмена и анизотропии. Ранее эта модель применялась для описания двумерных изотропных антиферромагнетиков в окрестности квантовых критических точек [46]; ее классический аналог использовался для оценки величины температур Кюри/Нееля изотропных классических магнетиков в пространстве размерности *d*=2+ε и определения соответствующих критических индексов [62]. Как показано ниже, в случае квантовых магнетиков величина температуры Кюри/Нееля может быть выражена через параметры основного состояния с хорошей точностью несмотря на континуальный характер модели (2.52), поскольку лишь возбуждения с волновым вектором $q < (T/J)^{1/2} \ll 1$ играют при этом существенную роль (см. раздел 2.2).

В отсутствие анизотропии и внешнего поля модель (2.52) обладает $O(3)/O(2)$ группой симметрии, соответствующей возможности вращения одного из трех базисных векторов в трехмерном пространстве ($O(2)$ - группа операций вращения в базисной плоскости, оставляющих неизменным выбранный базисный вектор). В отличие от случая квантового ферромагнетика, модель (2.52) может быть обобщена на случай *N*-мерного спинового пространства $O(N)/O(N-1)$ с произвольным $N$ путем введения поля $\sigma_i = \{\sigma_1...\sigma_N\}$ и замены $\sigma^z$ на $\sigma_N$.

## 2.4. Описание двумерного температурного режима в рамках ренормгруппового подхода

В «двумерном гейзенберговском» режиме взаимодействие спиновых волн является существенным, но сами спин-волновые возбуждения являются хорошо определенными. Наличие такого режима является специфической особенностью квазидвумерных систем с малыми значениями межплоскостного обмена и анизотропии. Как можно увидеть уже из результатов спин-волновых подходов (раздел 2.2), в этом режиме имеются логарифмические расходимости в (подрешеточной) намагниченности, определяемые параметрами $\ln(\xi_{J'}/L_\tau)$ в квантовом и $\ln(\xi_{J'}/a)$ в классическом случае.

Чтобы выйти за рамки теории возмущений и определить эволюцию термодинамических свойств с увеличением температуры, необходимо выполнить суммирование этих расходимостей. Наиболее эффективным инструментом здесь является ренормгрупповой (РГ) подход [46,60-63], вводящий перенормированную модель, которая не содержит логарифмических расходимостей и рассматривающий эволюцию ее параметров с изменением масштабов длины. Эта модель допускает описание в рамках теории возмущений (в нашем случае - спин-волновой теории).

Рассмотрим вначале случай квантового антиферромагнетика. В этом случае в качестве эффективных параметров следует рассмотреть подрешеточную намагниченность и температуру фазового перехода; эти величины удобно выражать через наблюдаемые параметры основного состояния: намагниченность $\overline{S}_0$, спиновую жесткость $\rho_s$, скорость спиновых волн $c$, межплоскостной обмен $\alpha_r$ и анизотропию $f_r = (\Delta/\rho_s)\overline{S}_0$ ($\Delta$ – щель в энергетическом спектре). На первом шаге ренормгруппового преобразования вводятся параметры квантовой перенормировки $Z_i$ согласно соотношениям

$$\bar{S}_0 = ZS, g_0 = gZ_1, c_0 = cZ_c,$$
$$f = f_r Z_2, \alpha = \alpha_r Z_3,$$

связывающим наблюдаемые параметры основного состояния $g, c, \alpha_r, f_r$ с (затравочными) параметрами исходной модели $g_0, c_0, \alpha, f$, где $g = c/\rho_s$ и $g_0 = c_0/\rho_s^0$ – безразмерные перенормированная и затравочная константа связи модели (2.52).

В силу неуниверсальности перенормировочных констант $Z_i$, то есть их зависимости от деталей структуры решетки, они могут быть определены лишь из рассмотрения исходной решеточной (неконтинуальной) версии производящего функционала (2.47). Указанные параметры могут быть вычислены в спин-волновой теории, являющейся фактически разложением в ряд по $g$ ($g \sim 1/S$ для больших $S$). Для антиферромагнетиков с квадратной решеткой результаты раздела 2.2 приводят к выражениям

$$Z = 1/Z_1 = Z_2 = Z_3^{1/2} = 1 - 0.197/S, \qquad (2.53)$$
$$Z_c = 1 + 0.079/S$$

с точностью до членов первого порядка по $1/S$ [16,38-45]. Для учета квантовых перенормировок удобно иметь эквивалент результатов (2.53), определенный в рамках континуальной модели (2.52). В первом порядке по $g$ находим

$$Z = 1 - (N-1)\frac{g\Lambda}{4\pi} + O(g^2),$$
$$Z_1 = 1 - (N-2)\frac{g\Lambda}{4\pi} + O(g^2), Z_c = 1 + O(g^2), \qquad (2.54)$$
$$Z_2 = 1 + \frac{g\Lambda}{2\pi} + O(g^2), Z_3 = 1 + \frac{3g\Lambda}{4\pi} + O(g^2),$$

где $\Lambda$ – ультрафиолетовый параметр обрезания, необходимый для регуляризации расходимостей, возникающих при вычислении параметров основного состояния. После выполнения квантовой перенормировки (2.54) теория, как мы увидим ниже, становится полностью универсальной, так что термодинамические свойства не зависят от параметра обрезания $\Lambda$.

Результирующая теория, однако, содержит логарифмические расходимости $\ln(\xi_{J'}/L_\tau)$, являющиеся следствием двумерного характера спиновых флуктуаций в рассматриваемом температурном режиме. Для суммирования указанных расходимостей в рамках РГ подхода на втором этапе РГ преобразования вводится формальный инфракрасный параметр обрезания $\mu$, так что указанные расходимости заменяются на $\ln[1/(\mu L_\tau)]$. Далее рассматриваются температурно-зависящие перенормировочные параметры $\tilde{Z}_i$, введенные согласно теоретико-полевой формулировке РГ [61,67]

$$g = g_R \tilde{Z}_1, u_r = u_R \tilde{Z}_u,$$
$$\pi = \pi_R \tilde{Z}, h = h_R \tilde{Z}_1 / \sqrt{\tilde{Z}},$$
$$f_r = f_R \tilde{Z}_2, \alpha_r = \alpha_R \tilde{Z}_3. \tag{2.55}$$

являющиеся функциями $\mu$ и определяющимися из условия отсутствия логарифмических расходимостей в перенормированной теории; индекс $R$ соответствует квантово- и температурно-перенормированным величинам; $u_r = c/T$. Аналогично классической нелинейной сигма-модели [61], введение пяти перенормировочных параметров для пяти независимых параметров модели оказывается достаточным для того, чтобы устранить все имеющиеся расходимости (см. также [67]).

Бесконечно малое изменение $\mu$ генерирует преобразование ренормгруппы. Для наших целей необходимо точно учесть сингулярные вклады в диаграммах, которые содержат две петли, образуемые линиями функций Грина поля $\pi$ (так называемое «двухпетлевое» приближение). Соответствующий результат для эффективной температуры $t = g/(4\pi u) = T/(4\pi \rho_s)$ и подрешеточной намагниченности в перенормированной модели имеет вид

$$\mu \frac{dt_r}{d\mu} = -(N-2)t_r^2 - (N-2)t_r^3 + O(t_r^4), \qquad (2.56)$$

$$\mu \frac{d\ln Z}{d\mu} = \frac{1}{2}(N-1)t_r + O(t_r^3). \qquad (2.57)$$

В перенормировке параметров межплоскостного обмена и анизотропии достаточно рассмотреть сингулярные вклады диаграмм с одной петлей, что приводит к результату

$$\mu \frac{d\ln Z_2}{d\mu} = -2t_r + O(t_r^2), \qquad (2.58)$$

$$\mu \frac{d\ln Z_3}{d\mu} = -t_r + O(t_r^2). \qquad (2.59)$$

Уравнения (2.56) – (2.59) определяют эволюцию параметров модели при РГ преобразовании с изменением параметра масштаба $\mu$, причем конечное значение параметра $\mu \sim \max(\alpha, f^{1/2})$ определяется условием отсутствия логарифмических расходимостей в эффективной модели.

Решение уравнений (2.58), (2.59) позволяет получить температурную перенормировку параметров межплоскостного обмена и анизотропии:

$$f_t/f_r = \overline{\sigma}_r^{4/(N-1)}\left[1 + O(t_r/\overline{\sigma}_r^{1/\beta_2})\right], \qquad (2.60)$$

$$\alpha_t/\alpha_r = \overline{\sigma}_r^{2/(N-1)}\left[1 + O(t_r/\overline{\sigma}_r^{1/\beta_2})\right]. \qquad (2.61)$$

где $\overline{\sigma}_r$ - относительная подрешеточная намагниченность. Последняя определяется решением уравнений (2.56), (2.57), которое имеет вид [63]:

$$\overline{\sigma}_r^{1/\beta_2} = 1 - \frac{t_r}{2}\left[(N-2)\ln\frac{2}{u_r^2 \Delta(f_t, \alpha_t)} + \frac{2}{\beta_2}\ln(1/\overline{\sigma}_r) + 2(1 - \overline{\sigma}_r^{1/\beta_2}) + O(t_r/\overline{\sigma}_r^{1/\beta_2})\right], \qquad (2.62)$$

где

$$\Delta(f, \alpha) = f + \alpha + \sqrt{f^2 + 2\alpha f}, \qquad (2.63)$$

универсальная функция анизотропии и межплоскостного обмена. Величина

$$\beta_2 = \frac{N-1}{2(N-2)} \qquad (2.64)$$

является пределом критического индекса намагниченности $\beta_{2+\varepsilon}$ в пространстве размерности $d = 2 + \varepsilon$ [61] при $\varepsilon \to 0$; в физически важном случае $N=3$ имеем $\beta_2=1$. Ведущий логарифмический член в квадратных скобках (2.62) соответствует результату ССВТ (2.31), в то время как другие два члена описывают поправки к этой теории. Результаты (2.60)-(2.62) при $N=3$ совпадают с результатами ССВТ (2.23) и (2.24).

Как уже указывалась, температура Нееля не может быть вычислена непосредственно в РГ подходе, поскольку возбуждения неспинволнового типа дают вклад вблизи $T_M$, что приводит к необходимости учета диаграмм с произвольным числом петель. Однако, можно получить общее выражение для температуры Нееля следующим образом. Рассмотрим сначала температуру $t_r^*$, определяющую переход (кроссовер) к критическому режиму. В рамках РГ анализа она определяется условием перехода в режим сильной связи рассмотренных уравнений РГ, т.е. условием $t_\rho \sim 1$ (что эквивалентно $t_r^* \sim \bar{\sigma}_r^{1/\beta_2}$). При дальнейшим РГ преобразовании трехмерные гейзенберговские (или двумерные изинговские) флуктуации могут изменить только постоянный (несингулярный) член $O(t_r/\bar{\sigma}_r^{1/\beta_2})$, который заменяется универсальной функцией $\Phi_{\text{AF}}(\alpha_r/f_r) \sim 1$. Таким образом, для температуры Нееля находим

$$t_{\text{N}} = 2\left[(N-2)\ln\frac{2}{u_r^2 \Delta(f_c,\alpha_c)} + 2\ln(2/t_{\text{N}}) + \Phi_{\text{AF}}(\alpha_r/f_r)\right]^{-1}, \qquad (2.65)$$

где

$$f_c = f_r t_{\text{N}}^{2/(N-2)},\ \alpha_c = \alpha_r t_{\text{N}}^{1/(N-2)} \qquad (2.66)$$

[$t = T/(2\pi\rho_s), u_r = c/T$]. Второй член в знаменателе (2.65), представляющий собой поправку к ССВТ, имеет порядок $\ln\ln(2T_{\text{N}}^2/\alpha)$ и приводит к существенному понижению температуры Нееля по сравнению с ее значением в ССВТ. Функция Ф определяется неспинволновыми возбуждениями и не может быть вычислена в рамках РГ подхода. В квазидвумерном изотропном случае Ф может быть вычислена с помощью 1/$N$ разложения (см. следующий раздел); более общий случай требует численного анализа (например, квантовым методом Монте-Карло).

Ферромагнитный случай может быть рассмотрен способом, аналогичным антиферромагнитному. В этом случае перенормировки, обусловленные квантовыми флуктуациями, отсутствуют, так что достаточно ограничиться температурными перенормировками. Чтобы построить теорию возмущений для производящего функционала (2.47), удобно перейти от вещественных полей $\pi_x, \pi_y$ к циклическим компонентам

$$\pi^\pm = \pi_x \pm i\pi_y. \qquad (2.67)$$

Затравочные функции Грина полей $\pi^+, \pi^-$ имеют вид

$$G^{(0)}(\mathbf{k}, i\omega_n) = \frac{1}{g}\left[i\omega_n + k_\parallel^2 + \alpha(1 - \cos k_z) + f + h\right]^{-1}, \qquad (2.68)$$

где $g = 1/S$. Поскольку квантовая перенормировка отсутствует, $Z_i \equiv 1$, индексы $r$ у параметров основного состояния могут быть опущены. Множители $\tilde{Z}_i$ имеют тот же самый вид, как и в антиферромагнитном случае с $N = 3$ и заменой $\ln(u_r\mu) \to \ln(u\mu)/2$, где $u = JS/T$. Для относительной намагниченности $\bar{\sigma} \equiv \bar{S}/S$ ($\bar{S} = \langle S^z \rangle$) и температуры Кюри получаем уравнения (2.62) и (2.65) с заменами $u_r^2 \to u, t_r \to t$ [63].

В классическом случае затравочная функция Грина поля $\pi = \mathbf{n} - (\mathbf{nz})\mathbf{z}$ имеет вид

$$G^{(0)}(\mathbf{k}) = \frac{1}{2\pi t}[2(2 - \cos k_x a - \cos k_y a) + \alpha(1 - \cos k_z a) + f + h]^{-1} \qquad (2.69)$$

где $a$ - постоянная решетки. Выражение для намагниченности содержит неуниверсальные вклады, определяемые конкретной структурой решетки. Перенормировочные константы в этом случае могут быть представлены как

$$Z_i(t, a) = Z_{Li}(t)\tilde{Z}_i(t_L, a) \qquad (2.70)$$

где $t_L = tZ_{L1}^{-1}$, $Z_{Li}$ содержат нелогарифмические вклады, которые не изменяются при РГ преобразованиях, и определяются из теории возмущений, $\tilde{Z}_i$ содержат все остальные вклады. Уравнение для намагниченности в двухпетлевом РГ подходе для классического магнетика может быть получено как и для квантового случая, при этом в результатах (2.62) и (2.65) для $\bar{\sigma}$ и $T_\text{M}$ необходимо заменить $u_r^2 \to 1/64, t_r \to t_L$ [63].

Таким образом, РГ подход достаточен для вычисления намагниченности при температурах, не слишком близких к температуре магнитного перехода, при которых спин-волновые возбуждения играют решающую роль, а также позволяет вычислить температуры Кюри (Нееля) с точностью до некоторый постоянной, являющейся универсальной в квантовом случае.

## 2.5. Описание критического режима и вычисление температуры Нееля квантовых квазидвумерных антиферромагнетиков

Учет неспинволновых возбуждений, необходимых для правильного описания критической области и полного вычисления температуры Нееля, возможен в рамках $1/N$ разложения. Это разложение основано на упрощении исходной модели в пределе $N \to \infty$ и было впервые использовано при вычислении критических индексов в модели $\phi^4$ (см., напр., [68]). Позже $1/N$ разложение успешно применялось для описания свойств квантовой модели Гейзенберга вблизи квантовой критической точки [27].

Для модели Гейзенберга предел $N \to \infty$ совпадает с так называемой сферической моделью [64], пренебрегающей связью различных спиновых компонент. При этом физическое условие $\mathbf{S}_i^2 = S(S+1)$ заменяется условием на среднее по узлам значение:

$$\sum_i \mathbf{S}_i^2 = NS(S+1), \qquad (2.71)$$

Такое приближение приводит к резкому упрощению модели, позволяя решить ее точно. Дальнейшие поправки, вычисляемые путем разложения в окрестности седловой точки, дают последовательное улучшение приближения (2.71) по параметру $1/N$.

Рассмотрим применение $1/N$ разложения к вычислению температур Кюри (Нееля) квазидвумерных магнетиков [28]. В квантовом случае вновь используется обобщение модели Гейзенберга на модель с $O(N)/O(N-1)$ симметрией (2.52). Рассмотрим вначале изотропный случай. В связи с наличием дальнего порядка ниже температуры Нееля, произведем сдвиг поля $\sigma = \tilde{\sigma} + \bar{\sigma}$ где $\bar{\sigma}$ – относительная подрешеточная намагниченность $\bar{S}/S$. Диагональные элементы функций Грина поля $\tilde{\sigma}$

$$G^{mn}(\mathbf{k}, q_z, \omega_n) = \rho_s^0 \int d\tau \langle T[\tilde{\sigma}_{\mathbf{k},q_z}^m(\tau)\tilde{\sigma}_{-\mathbf{k},-q_z}^n(0)] \rangle \tag{2.72}$$

(которые только и отличны от нуля) пропорциональны неоднородной динамической спиновой восприимчивости:

$$G^{mn}(\mathbf{k}, k_z, \omega) = \frac{\rho_s^0}{S^2} \chi^{mm}(\mathbf{k} + \mathbf{Q}, k_z + \pi, \omega)\delta_{mn}, \tag{2.73}$$

где $\mathbf{Q} = (\pi, \pi)$ – волновой вектор антиферромагнитной структуры в плоскости; для $N = 3$

$$\chi^{\alpha\beta}(\mathbf{k}, k_z, \omega) = \sum_i \exp[i(\mathbf{k}\mathbf{R}_i + k_z R_i^z)]\langle\langle S_0^\alpha | S_i^\beta \rangle\rangle_\omega, \tag{2.74}$$

$S_i^\alpha$ – операторы спина, $\alpha, \beta = x, y, z$. Для определенности далее полагаем, что подрешеточная намагниченность направлена в $N$-м направлении, т.е. $\bar{\sigma}^m = \bar{\sigma}\delta_{mN}$. Тогда $G^{NN}$ соответствует продольной функции Грина $G_l$, в то время как другие диагональные компоненты – поперечной функции Грина $G_t$. Условие $\sigma_i^2 = 1$ в указанных обозначениях принимает вид

$$1 - \bar{\sigma}^2 = \frac{T}{\rho_s^0} \sum_{\omega_n} \sum_m \int \frac{d^2\mathbf{k}}{(2\pi)^2} \int \frac{dk_z}{2\pi} G^{mm}(k, k_z, \omega_n). \tag{2.75}$$

После интегрирования по $\tilde{\sigma}$ производящий функционал (2.52) принимает вид

$$Z[h] = \int D\lambda \exp(NS_{eff}[\lambda, h]) \tag{2.76}$$

$$S_{eff}[\lambda, h] = \frac{1}{2}\ln\det\hat{G}_0 + \frac{1}{2g}(1 - \bar{\sigma}^2)\mathrm{Sp}(i\lambda)$$

$$+\frac{1}{2g}\mathrm{Sp}\left[\left(i\lambda\bar{\sigma}-h/\rho_s^0\right)\hat{G}_0\left(i\lambda\bar{\sigma}-h/\rho_s^0\right)\right], \tag{2.77}$$

где

$$\hat{G}_0 = [\partial_\tau^2/c_0^2 + \nabla^2 + \alpha\Delta_z]^{-1}, \tag{2.78}$$
$$\Delta_z \sigma_{i_z}(\mathbf{r},\tau) = \sigma_{i_z+1}(\mathbf{r},\tau) - \sigma_{i_z}(\mathbf{r},\tau).$$

Поскольку $N$ входит в действие (2.77) лишь как множитель в показателе экспоненты, предел $N \to \infty$ [сферическая модель (2.71)] соответствует приближению седловой точки функционала $S_{eff}[\lambda, h]$, в котором пренебрегается флуктуациями поля $\lambda$. При $T < T_N$ седловая точка имеет координаты $i\lambda = 0$ и $\bar{\sigma}^2 \neq 0$. При этом для спиновой функции Грина имеем

$$G_0(k, k_z, \omega_n) = \left[\omega_n^2 + k^2 + \alpha(1-\cos k_z)\right]^{-1}. \tag{2.79}$$

Температура Нееля, определенная из (2.75), равна

$$T_N^0 = \frac{4\pi\rho_s^{N=\infty}}{N\ln(2T_N^2/\alpha c^2)}, \tag{2.80}$$

где $\rho_s^{N=\infty} = N(1/g - 1/g_c)$ – жесткость спиновых волн в нулевом порядке по $1/N$, $g_c = 2\pi^2/\Lambda$ – формальный параметр теории. Значение (2.80) в $N/(N-2)$ раз ниже результата ССВТ (2.34) и РГ подхода (2.62). Это отличие обусловлено недостатком приближения сферической модели, рассматривающей различные спиновые компоненты независимо друг от друга.

В первом порядке по $1/N$ учитываются наинизшие поправки к условию (2.71), обусловленные однократным обменом возбуждением поля $\lambda$, которое учитывает связь между различными компонентами спина на узле. Общий вид уравнения для намагниченности при $T \gg \alpha^{1/2}$ и $\ln(2T_N^2/\alpha c^2) \gg 1$ в первом порядке по $1/N$ приведен в работе [28]. В области температур не слишком близких к точке магнитного перехода, где

$$NT/4\pi\rho_s \ll \bar{\sigma}^2/\bar{\sigma}_0^2, \tag{2.81}$$

уравнение для намагниченности имеет вид

$$(\bar{\sigma}/\bar{\sigma}_0)^{1/\beta_2}\left[1 - I_2(x_{\bar{\sigma}})\right]$$

$$= 1 - \frac{NT}{4\pi\rho_s}\left[(1-\frac{2}{N})\ln\frac{2T^2}{\alpha_r} + \frac{3}{N}\ln\frac{\overline{\sigma}_0^2}{\overline{\sigma}^2} - \frac{2}{N}\frac{\ln(2T^2/\alpha_r)}{\ln(2T^2/\alpha_r)+x_{\overline{\sigma}}} - I_1(x_{\overline{\sigma}})\right]. \quad (2.82)$$

Здесь $I_{1,2}(x_{\overline{\sigma}})$ – некоторые функции (см. [28]),

$$x_{\overline{\sigma}} = \frac{4\pi\rho_s}{(N-2)T}\frac{\overline{\sigma}^2}{\overline{\sigma}_0^2}. \quad (2.83)$$

$\overline{\sigma}_0$ и $\rho_s$ – подрешеточная намагниченность и спиновая жесткость основного состояния в нелинейной сигма-модели квантового двумерного антиферромагнетика [27],

$$\frac{\overline{\sigma}_0^2}{\rho_s} = \frac{g}{N}\left(1 - \frac{8}{3\pi^2 N}\ln\frac{N\Lambda}{16\rho_s}\right), \quad (2.84)$$

$$\rho_s = \rho_s^{N=\infty}\left(1 + \frac{32}{3\pi^2 N}\ln\frac{N\Lambda}{16\rho_s}\right), \quad (2.85)$$

$\alpha_r$ – перенормированный параметр межплоскостного обмена,

$$\alpha_r = \alpha\left(1 - \frac{8}{3\pi^2 N}\ln\frac{N\Lambda}{16\rho_s}\right). \quad (2.86)$$

Как и в ренормгрупповом подходе, намагниченность подрешетки, выраженная в терминах квантово-перенормированнных величин $\rho_s, \overline{\sigma}_0$ и $\alpha$, не зависит от параметра обрезания $\Lambda$, то есть является универсальной величиной. Результат (2.82) отличается от результата ренормгруппы (2.62) лишь коэффициентом при субведущем члене $\ln(\overline{\sigma}_0/\overline{\sigma})$ (6/*N* вместо 3/β$_2$), что лежит за пределами точности первого порядка по *1/N*. Давая качественно правильное описание двумерного режима, уравнение (2.82) позволяет описать подрешеточную намагниченность и в режиме, переходном к критическому.

В критическом режиме $\overline{\sigma}$ достаточно мало, чтобы удовлетворить условию

$$\overline{\sigma}^2/\overline{\sigma}_0^2 \ll (N-2)T/4\pi\rho_s, \quad (2.87)$$

при котором $x_{\overline{\sigma}} \ll 1$. В этом режиме результат 1/*N* разложения для подрешеточной намагниченности имеет вид

$$\frac{\overline{\sigma}}{\overline{\sigma}_0} = \left[\frac{4\pi\rho_s}{(N-2)T_N}\right]^{(\beta_3/\beta_2-1)/2} \left[\frac{1}{1-A_0\left(1-\frac{T}{T_N}\right)}\right]^{\beta_3}, \qquad (2.88)$$

где

$$\beta_3 = \frac{1}{2}\left(1-\frac{8}{\pi^2 N}\right) \qquad (2.89)$$

– критический индекс намагниченности, $A_0 = 2.8906/N$. При $N = 3$ имеем $\beta_3 \simeq 0.36$, что совпадает с результатом $1/N$ разложения в модели $\phi^4$ [68] при $d = 3$, в согласии с гипотезой универсальности. Уравнение для температуры Нееля $T_N$ имеет вид

$$T_N = 4\pi\rho_s \left[(N-2)\ln\frac{2T_N^2}{\alpha_r} + 3\ln\frac{4\pi\rho_s}{(N-2)T_N} - 0.0660\right]^{-1}. \qquad (2.90)$$

Результаты для намагниченности в низкотемпературной (2.82) и критической (2.88) области плавно сшиваются друг с другом, см. Раздел 2.6.

Спектр возбуждений в точке магнитного фазового перехода определяется собственно-энергетической частью $\Sigma(k,k_z,0)$ при $T = T_N$. При $\alpha^{1/2} \ll k \ll T_N$ соответствующее выражение для функции Грина имеет вид ($G = G_t = G_l$)

$$G(k,k_z,0) = \frac{1}{k^2}\left[\frac{(N-2)T_N}{4\pi\rho_s}\ln\frac{2k^2}{\alpha}\right]^{1/(N-2)} \frac{N-1}{N}\left[1 - \eta\ln\frac{N\Lambda}{16\rho_s}\right]. \qquad (2.91)$$

Нетривиальная логарифмическая зависимость $G(k,k_z,0)$ отличает этот результат от ранее рассмотренных результатов спин-волновой теории. Множитель $N–1$ в (2.91) имеет простой физический смысл – это число голдстоуновских мод в $N$-компонентной модели. При $k \ll \alpha^{1/2}$, $k_z \ll 1$ импульсная зависимость функции Грина изменяется:

$$G^{-1}(k,k_z,0) = (1+A_1)\alpha_c^{\eta/2}\left(k^2 + \frac{\alpha_c}{2}k_z^2\right)^{1-\eta/2} \quad (k \ll \alpha^{1/2}, k_z \ll 1), \qquad (2.92)$$

$$A_1 = \eta\ln\frac{N\Lambda}{16\rho_s} + \frac{1}{N}\ln\ln\frac{2T^2}{\alpha} + \frac{0.4564}{N}, \quad A_2 = -0.6122/N \qquad (2.93)$$

где

$$\eta = 8/(3\pi^2 N) \qquad (2.94)$$

– трехмерный критический индекс для асимптотики корреляционной функции в точке перехода в первом порядке по $1/N$. В этом режиме, как видно из (2.92), возбуждения имеют трехмерный характер и характеризуются критическим индексом $\eta$ ($\eta \simeq 0.09$ при $N = 3$). Остальные критические индексы могут быть найдены из скейлинговых соотношений, которые сохраняются в рамках регулярного $1/N$ разложения [68]. Величина

$$\alpha_c = \alpha(1 + A_2)/(1 + A_1) \qquad (2.95)$$

может интерпретироваться как перенормированный параметр межплоскостного обмена при $T = T_N$.

С учетом (2.95) находим соотношение между перенормированными параметрами обмена при низких $T$ и при $T = T_N$:

$$\alpha_c = \alpha_r \left(1 + \frac{1.0686}{N}\right) \left[\frac{(N-2)T_N}{4\pi\rho_s}\right]^{1/(N-2)}. \qquad (2.96)$$

Так же как в ССВТ (см. раздел 2.2), перенормированное значение параметра межплоскостного обмена в $T_N$ меньше, чем его низкотемпературное значение, но конкретное выражение при $N = 3$ отличается от результата ССВТ численным множителем, примерно равным 1.3.

При наличии малой анизотропии типа «легкая ось» спектр возбуждений в нулевом порядке по $1/N$, определяемый полюсами невозмущенной продольной и поперечной функций Грина, содержит щель $\Delta = \Delta(\alpha_r, f_r)$ для всех компонент $\sigma_m$, кроме $m = N$:

$$\begin{aligned} G_t^0(k, \omega_n) &= \left[\mathbf{k}^2 + \omega_n^2 + 2\alpha(1 - \cos k_z) + \Delta\right]^{-1}, \\ G_l^0(k, \omega_n) &= \left[\mathbf{k}^2 + \omega_n^2 + 2\alpha(1 - \cos k_z)\right]^{-1}. \end{aligned} \qquad (2.97)$$

При достаточно низких температурах

$$T \gg f_r^{1/2}, \; NT \ln(T^2/f_r)/4\pi\rho_s \ll \bar{\sigma}^2/\bar{\sigma}_0^2 \qquad (2.98)$$

$1/N$ разложение воспроизводит результат теории спиновых волн (2.39). При промежуточных температурах

$$TN\ln(T^2/f_r)/4\pi\rho_s \gg \overline{\sigma}^2/\overline{\sigma}_0^2 \gg TN/4\pi\rho_s \qquad (2.99)$$

имеем

$$\left(\overline{\sigma}/\overline{\sigma}_0\right)^{1/\beta_2} = 1 - \frac{T}{4\pi\rho_s}\left[(N-2)\ln\frac{T^2}{f_r c^2} + B_2\ln\frac{\overline{\sigma}_0^2}{\overline{\sigma}^2} - 2 + 2\frac{\overline{\sigma}^2}{\overline{\sigma}_0^2} + O(\frac{NT}{4\pi\rho_s}\frac{\overline{\sigma}_0^2}{\overline{\sigma}^2})\right] \qquad (2.100)$$

где

$$B_2 = 3 + f_r/\sqrt{f_r^2 + 2\alpha_r f_r} \qquad (2.101)$$

Температурно-зависимый параметр анизотропии может быть определен из выражения

$$[G_t(k,0)]^{-1} \equiv [G_t^0(k,0)]^{-1} + \Sigma_t(k,0) - \Sigma_l(0,0) \propto k^2 + f(T), \qquad (2.102)$$

откуда находим

$$f(T)/f_r = (\overline{\sigma}/\overline{\sigma}_0)^{4/(N-2)}. \qquad (2.103)$$

В критической области, определяемой неравенством $\overline{\sigma}^2/\overline{\sigma}_0^2 \ll TN/4\pi\rho_s$ имеет место поведение намагниченности изинговского типа и $1/N$ – разложение не применимо. Это проявляется, в частности, в том, что производная $\partial\overline{\sigma}/\partial T$ расходится при некоторой температуре $\tilde{T}_M$. Однако соответствующая критическая область очень узка (как и критическая область для трехмерных флуктуаций в изотропном квазидвумерном случае). Поэтому температура Нееля может быть оценена как $T_N \simeq \tilde{T}_M$.

Суммируем результаты в практически важном случае $N=3$. В спин-волновой и двумерной областях, то есть при

$$\overline{\sigma}_r \gg T/4\pi\rho_s, \Gamma \gg \Delta \qquad (2.104)$$

результат РГ для относительной (подрешеточной) намагниченности имеет вид

$$\overline{\sigma}_r = 1 - \frac{T}{4\pi\rho_s}\left[\ln\frac{2\Gamma(T)}{\Delta(f_t,\alpha_t)} + 2\ln(1/\overline{\sigma}_r) + 2(1-\overline{\sigma}_r)\right], \qquad (2.105)$$

где функция $\Delta(f,\alpha)$ определена в (2.63), температурно-перенормированные значения межплоскостного обмена и параметра анизотропии

$$f_t/f_r = (\alpha_t/\alpha_r)^2 = \overline{\sigma}_r^2 \qquad (2.106)$$

и величины $\Gamma(T), \bar{\sigma}_r, f_r, \alpha_r, \rho_s$ определены в Таблице 1. Уравнение для $T_M$ имеет вид

$$T_M = 4\pi\rho_s / \left[ \ln\frac{2\Gamma(T_M)}{\Delta(f_c,\alpha_c)} + 2\ln\frac{4\pi\rho_s}{T_M} + \Phi(f/\alpha) \right], \qquad (2.107)$$

где $\Phi(x)$ – некоторая (универсальная в квантовом случае) функция порядка единицы, $f_c$ и $\alpha_c$ – параметры межплоскостного обмена и анизотропии при $T = T_M$, причем

$$f_c/f_r = (\alpha_c/\alpha_r)^2 = (T_M/4\pi\rho_s)^2. \qquad (2.108)$$

Так как $T_M/4\pi\rho_s \sim 1/\ln(1/\Delta) \ll 1$, температурные перенормировки важны для правильного описания экспериментальных данных. В частности, параметры межплоскостного обмена и анизотропии, измеренные при различных температурах, могут значительно отличаться. Результаты ССВТ (2.31) – (2.40) в пределе нулевой анизотропии (межплоскостного обмена) отличаются от (2.105) заменой $4(3) \to 2(1)$ для коэффициента во втором члене в квадратных скобках. Таким образом, роль поправок к ССВТ более важна в изотропном квазидвумерном магнетике, чем в двумерном анизотропном.

Результат $1/N$ разложения в $O(N)$ модели вне критической области, точнее при

$$\bar{\sigma}_r^2 > T/4\pi\rho_s, \Gamma \gg \Delta \qquad (2.109)$$

в первом порядке по $1/N$ имеет вид

$$\bar{\sigma}_r = 1 - \frac{T}{4\pi\rho_s}\left[ \ln\frac{2\Gamma(T)}{\Delta(f_t,\alpha_t)} + 2B_2\ln(1/\bar{\sigma}_r) + 2(1-\bar{\sigma}_r^2) \right]. \qquad (2.110)$$

В частных случаях $\alpha = 0$ и $f = 0$ коэффициент при втором члене в квадратных скобках в (2.110) вдвое больше чем для РГ результата (2.105), что обеспечивает более правильное описание температурной области, переходной к критическому поведению, и критической области. Уравнения для $T_M$ имеют вид (2.107), одинаковый в обоих подходах. В изотропном случае (*f* = 0) результат 1/*N* разложения для подрешеточной намагниченности в критической области

$$\bar{\sigma}_r = \left(\frac{4\pi\rho_s}{T_N}\right)^{(\beta_3-1)/2} \left[\frac{1}{1-A_0}\left(1-\frac{T}{T_N}\right)\right]^{\beta_3}, \qquad (2.111)$$

где $A_0 = 0.9635$ и $\beta_3 \simeq 0.36$.

## 2.6. Теоретическое описание экспериментальных данных намагниченности и температур Нееля слоистых систем

Рассмотрим теперь применение полученных результатов для анализа экспериментальных данных. Одним из хорошо изученных слоистых соединений является $La_2CuO_4$ [9,69]. Значение перенормированного параметра обмена для этого соединения, $\gamma|J|\simeq 1850\,K$ может быть определено из экспериментальных данных для спин-волнового спектра при низких температурах [70], в то время как значение межплоскостного обмена $\alpha_r = 1\cdot 10^{-3}$ может быть найдено из сравнения намагниченности в ССВТ с экспериментальной зависимостью $\bar{\sigma}_r(T)$ при низких температурах [28,63]. На Рис. 4 представлены экспериментальная температурная зависимость намагниченности подрешетки в $La_2CuO_4$ [69], результаты спин-волновых приближений (СВТ, ССВТ и теории Тябликова) для этого соединения, РГ подхода и $1/N$ разложения. Результат для температуры Нееля $1/N$ разложения первого порядка – $T_N = 345\,K$, что значительно ниже всех спин-волновых приближений и находится в хорошем согласии с экспериментальным значением $T_N^{\exp} = 325\,K$.

РГ подход правильно описывает зависимость $\bar{\sigma}_r(T)$ в спин-волновой области $(T<300K)$ и области двумерных флуктуаций (которая очень узка при вышеприведенном малом значении $\alpha$), в то время как при более высоких температурах этот подход переоценивает $\bar{\sigma}$. С другой стороны, кривая $1/N$ разложения расположена ближе всего к экспериментальным данным и правильно описывает критическое поведение. Результаты численного решения уравнения в температурной области (2.109) и зависимости (2.111) в критической области, совпадают в точке $T = 330\,K$, отмеченной крестиком. Различие между теоретической и экспериментальной кривыми в температурной области $320\,K < T < 340\,K$ может быть обусловлено влиянием анизотропии. При фиксированном $\Delta$ в (2.110) и $B_2$, определенном из наилучшего совпадения с экспериментальными данными при промежуточных температурах (см. Рис. 4), находим значения $\alpha_r = 1 \cdot 10^{-4}$, $f_r = 5 \cdot 10^{-4}$. Таким образом, рассматриваемый подход дает возможность оценить относительную роль межплоскостного обмена и магнитной анизотропии в слоистых соединениях. Отметим, что альтернативное объяснение различия между теоретическим и экспериментальным результатами, основанное на рассмотрении циклического 4-х спинового взаимодействия, было предложено в работе [71].

В слоистых перовскитах $K_2NiF_4$, $Rb_2NiF_4$ и $K_2MnF_4$ магнитная анизотропия, как известно, является более важной, чем межплоскостной обмен. Соединение $K_2NiF_4$ имеет спин $S=1$, из данных нейтронного рассеяния следует $|J|=102 K$ и $T_N^{\exp} = 97.1 K$ [4]. На рис. 5 показана экспериментальная зависимость $\overline{\sigma}(T)$ [1] и результаты спин-волновых подходов, РГ подхода и численного решения уравнения (2.110). Значение $f_r = 0.0088$ было получено из сравнения результата намагниченности ССВТ с экспериментальными данными при низких температурах (это значение хорошо согласуется с экспериментальным $f_r = 0.0084$ [4]). В спин-волновом и двумерном флуктуационном температурных интервалах (2.104) ($T < 80 K$) кривые, соответствующие $1/N$ разложению и РГ подходу, располагаются несколько выше, чем экспериментальные точки, поскольку $T^2/f_r c^2$ в этой области не велико. В то же время кривая $1/N$ разложения находится в хорошем численном согласии с экспериментальными данными. Процедура экстраполяции к изинговскому критическому поведению дает $T_N = 91.4$ K, причем ширина критической изинговской области составляет 1K. Отметим, что учет членов порядка $1/x_{\overline{\sigma}}$ в (2.110) приводит к значению $T_N = 92.7 K$. В переходной к критическому поведению области 80K<$T$<90K теоретическая $O(3)$ кривая для $K_2NiF_4$, в отличие от случая $La_2CuO_4$ лежит слегка ниже экспериментальной. Этот факт может быть приписан влиянию межплоскостного обмена. Определение соответствующих параметров в переходной области приводит к значениям $\alpha_r = 0.0017$, $f_r = 0.0069$, которые соответствуют $T_{\text{Neel}} = 97\,\text{K}$ и затравочным параметрам $\alpha|J| = 0.1\,\text{K}$, $\zeta|J| = 0.76\,\text{K}$. Соответствующие экспериментальные данные для $\alpha$ отсутствуют, поэтому сравнение с экспериментом в данном случае затруднительно.

Соединение $Rb_2NiF_4$ обладает сильной магнитной анизотропией: согласно [4] $|J| = 82\,\text{K}$, $|J|f_r = 3.45\,\text{K}$, $T_N^{\text{exp}} = 94.5\,\text{K}$. Сравнение экспериментальной зависимости $\bar{\sigma}_r(T)$ с результатами ССВТ при низких температурах приводит к значению параметра анизотропии $f_r = 0.046$, в хорошем согласии с вышеприведенным экспериментальным значением. Из (2.107) следует $T_N = 95.5\,\text{K}$, что также находится близко к экспериментальным данным для температуры Нееля.

Соединение $K_2MnF_4$ имеет спин $S = 5/2$ и поэтому представляет собой промежуточную ситуацию между квантовым и классическим случаями. Параметры обмена и анизотропии $|J| = 8.4\,\text{K}$, $|J|f_r = 0.13\,\text{K}$ могут быть найдены из данных нейтронного рассеяния [4]. Рис. 6 показывает сравнение результатов различных подходов с экспериментальными данными для этого соединения. Можно видеть, что $1/N$ разложение приводит к результатам, хорошо описывающим экспериментальную ситуацию во всем интервале температур. В то же время экспериментальные точки расположены между квантовой и классической РГ кривыми, причем квантовое приближение является более удовлетворительным. Это подтверждает квантовый характер поправок к намагниченности даже при относительно большой величине спина. В рассматриваемом случае ССВТ, правильно учитывающая возбуждения на масштабе постоянной решетки, приводит к лучшим результатам по сравнению с РГ подходом. Таким образом, аккуратное рассмотрение систем с большим спином в рамках континуальных моделей требует численного расчета интегралов по импульсам и суммирования по мацубаровским частотам.

На Рис. 7 показано сравнение результатов ССВТ и РГ подхода для намагниченности классического магнетика с вычислениями методом Монте-Карло [72]. Можно видеть, что за исключением узкой критической области, кривая РГ довольно точна, хотя при этом пренебрегается топологическими возбуждениями. Область применимости РГ подхода в классическом случае более широка, чем в квантовом случае, так что нет необходимости использовать $1/N$ разложение для описания переходной и критической области.

Описанные результаты сравнения теоретических и экспериментальных данных по слоистым перовскитам суммированы в Таблице 2 и показывают, что РГ подход и $1/N$ разложение в $O(N)$ модели приводят к количественно правильным результатам температур магнитного перехода и намагниченности этих систем, находящихся в хорошем согласии с экспериментальными данными.

# 3. Квазидвумерные магнетики с анизотропией типа «легкая плоскость»

Другой важный класс низкоразмерных систем – двумерные системы с анизотропией типа «легкая плоскость». Классическая двумерная *XY* модель, соответствующая предельному случаю сильной легкоплоскостной анизотропии, была подробно изучена в ранних работах [73-75]. В указанных работах было продемонстрировано, что элементарными возбуждениями в этой модели являются топологические вихревые структуры и существует переход Березинского-Костерлица-Таулеса, связанный с диссоциацией вихревых пар при температуре

$$T_{BKT} = \frac{\pi}{2}|J|S^2. \tag{3.1}$$

При этой же температуре степенная зависимость корреляционной функции спинов от расстояния изменяется на экспоненциальную (в квантовой *XY* модели ситуация более сложная, поскольку должны быть учтены не только поперечные, но и продольные компоненты спина).

Более физически реальная ситуация, однако, описывается двумерной моделью Гейзенберга (2.1) со слабой анизотропией типа «легкая плоскость», т.е. $\eta, \zeta < 0$ и $|\eta|, |\zeta|, \alpha \ll 1$ (для удобства в дальнейшем сделаем замену $\eta \to -\eta$, $\zeta \to -\zeta$). В этом случае спин-волновые возбуждения при низких температурах играют определяющую роль в температурной зависимости (подрешеточной) намагниченности. Как и в случае «легкая ось», при температурах, не слишком низких по сравнению с температурой магнитного фазового перехода, необходим правильный учет динамического взаимодействия спиновых волн.

При слабой анизотропии «легкая плоскость», однако, переход Березинского-Костерлица-Таулеса предшествует магнитному фазовому переходу. При этом благодаря существованию «квазидальнего» порядка при $T < T_{BKT}$ включение сколь угодно слабого межплоскостного обмена приводит к появлению магнитного перехода выше $T_{BKT}$. Простое выражение для температуры Березинского-Костерлица-Таулеса, полученное в пределе малой анизотропии, имеет вид [76]

$$T_{BKT} = \frac{4\pi |J| S^2}{\ln(\pi^2/\eta)}. \qquad (3.2)$$

Как и для изотропных и легкоосных магнетиков, формула (3.2) недостаточна для количественного описания экспериментальных данных.

Аналогично магнетикам с анизотропией типа «легкая ось», можно ожидать, что термодинамические свойства этих систем, за исключением узкой окрестности $T_{BKT}$, определяются возбуждениями спин-волнового типа и для учета влияния взаимодействия спиновых волн при температурах вне критической области вновь может быть применен метод ренормгруппы [77].

РГ анализ снова выполняется на основе функционала (2.47). В классическом случае (то есть в пренебрежении динамической частью действия, содержащей производную по времени), имеется два типа возбуждений: поле $n_y$ описывает бесщелевые возбуждения в плоскости, а поле $n_z$ – возбуждения с поворотом спина поперек плоскости, обладающие щелью в энергетическом спектре. Разложение (2.47) по $n_{y,z}$ (ось квантования (подрешеточной) намагниченности предполагается вдоль $x$) приводит в ведущем порядке по $1/S$ к действию

$$L_{st} = \frac{1}{2} S^2 \int_0^{1/T} d\tau \sum_{\mathbf{k}} \left[ (J_0 - J_{\mathbf{k}}) \pi_{y\mathbf{k}} \pi_{y,-\mathbf{k}} + (J_0 - J_{\mathbf{k}} - \eta J_{\mathbf{k}}) \pi_{z\mathbf{k}+\mathbf{Q}} \pi_{z,-\mathbf{k}-\mathbf{Q}} \right], \qquad (3.3)$$

где $\mathbf{Q}$ – волновой вектор магнитной структуры и вектор $\boldsymbol{n}$ был представлен в виде $n_{\mathbf{k}}(\tau) = \{\sigma_{\mathbf{k}}(\tau), \pi_{y\mathbf{k}}(\tau), \pi_{z\mathbf{k}}(\tau)\}$.

При не слишком малых температурах возникают логарифмические вклады в (подрешеточную) намагниченность, суммирование которых является предметом РГ подхода. В отличие от случая «легкая ось», характерные энергетические масштабы возбуждений $\pi_y$ и $\pi_z$ различны. В связи с этим в намагниченности возникают два типа логарифмических вкладов: логарифмы анизотропии и логарифмы межплоскостного обмена. Ситуация, когда присутствуют два типа возбуждений с различными характерными масштабами, является типичной для систем, демонстрирующих переход (точнее, кроссовер) между двумя режимами с различными типами флуктуаций [67]. В рассматриваемой модели происходит переход от гейзенберговского (почти изотропного) поведения к $XY$ поведению.

Для правильного описания этого перехода необходимо включить анизотропию во все параметры перенормировки [67]. Из-за анизотропного характера модели имеется два параметра перенормировки поля $\pi$: $Z_{xy}$ и $Z_z$, определяемые с помощью соотношений $\pi_{xR}/\pi_x = \pi_{yR}/\pi_y = Z_{xy}$ и $\pi_{zR}/\pi_z = Z_z$.

Для этих параметров, а также эффективной температуры и анизотропии находим следующую систему РГ уравнений, определяющих изменение температуры, параметров анизотропии и межплоскостного обмена, а также параметра перенормировки поля π с изменением масштаба [77]

$$\mu \frac{d(1/t_\mu)}{d\mu} = (1+t_\mu)f(\eta_\mu,\mu) + O(t_\mu^2),$$

$$\mu \frac{d\ln Z_{xy}}{d\mu} = t_\mu \left[1 + f(\eta_\mu,\mu)\right] + O(t_\mu^3),$$

$$\mu \frac{d\ln \eta_\mu}{d\mu} = 2t_\mu f(\eta_\mu,\mu) + O(t_\mu^2),$$

$$\mu \frac{d\ln \alpha_\mu}{d\mu} = -t_\mu + O(t_\mu^2), \qquad (3.4)$$

где $\mu$ – параметр масштаба, $f(\eta_\mu,\mu) = \eta_\mu \mu^2/(\eta_\mu \mu^2 + \eta)$,

$$t = \begin{cases} T/(2\pi JS^2) & \text{ФМ} \\ T/(2\pi \rho_s) & \text{АФМ} \end{cases} \qquad (3.5)$$

– безразмерная температура, $\rho_s \simeq S(S+0.079)|J|$ – константа спиновой жесткости, $Z_z \equiv 1$.

Аналогично уравнениям (2.56) – (2.59), уравнения (3.4) описывают эволюцию параметров перенормированной модели с изменением масштаба. Выражение для эффективной температуры в этой модели имеет вид

$$\frac{1}{t_\mu} = \frac{1}{t} + \frac{1}{2}\ln\frac{\mu^2 t_\mu^2 + t^2\eta}{\mu_0^2 t_\mu^2 + t^2\eta} + \ln\frac{t}{t_\mu} + \Phi(\mu), \qquad (3.6)$$

где функция $\Phi(\mu) = O(t_\mu)$ соответствует вкладам более высокого порядка, $\mu_0 = q_0$ – начальный масштаб: $q_0 = \sqrt{T/JS}$ в ФМ случае, $q_0 = T/c$ в АФМ случае. В двумерном гейзенберговском режиме ($\mu \gg \sqrt{\eta}$) эффективная температура $t_\mu$ мала, так что

$$\frac{1}{t_\mu} = \frac{1}{t} + \ln\frac{\mu t}{\mu_0 t_\mu}, \qquad (3.7)$$

С другой стороны, при $\mu \ll \sqrt{\eta}$ взаимодействие спиновых волн приводит к изменению поведения $t_\mu$; находим

$$\frac{1}{t_\mu} = \frac{1}{t} - \ln\frac{\mu_0}{\sqrt{\eta}} + 2\ln\frac{t}{t_\mu} + \Phi(\mu). \tag{3.8}$$

В этом режиме $t_\mu$ зависит от $\mu$ только через функцию $\Phi(\mu)$. Величина $1/\eta^{1/2}$ есть характерный масштаб перехода (кроссовера) от гейзенберговского к $XY$ поведению, так что (3.8) описывает поведение эффективной температуры $t_\mu$ в $XY$ режиме. Аналогично температуре магнитного перехода магнетиков с анизотропией типа «легкая ось», температура Березинского-Костерлица-Таулеса может быть оценена из условия перехода в режим сильной связи, $t_\mu \sim 1$,

$$t_{BKT} = \left[\ln(\mu_0/\sqrt{\eta}) + 2\ln(2/t_{BKT}) + C\right]^{-1}, \tag{3.9}$$

где $C$ – универсальная постоянная.

Результат (3.9) может быть получен также из сравнения (3.8) с результатом решения РГ уравнений эффективной классической XY модели [74,75]. Действительно, даже если исходная модель – квантовая, на масштабах $\mu \ll \sqrt{\eta} \ll L_\tau^{-1}$ эффективная $XY$ модель является классической, поскольку $L_\tau$ определяет характерный масштаб, отделяющий квантовые флуктуации от классических. В связи с этим стандартная система РГ уравнений двумерной классической XY модели [73-75]

$$\mu\frac{d(1/t_\mu)}{d\mu} = 32\pi^2 y_\mu^2,$$
$$\mu\frac{dy_\mu}{d\mu} = -y_\mu(2 - \frac{1}{2t_\mu}) \tag{3.10}$$

может быть использована для описания РГ преобразования в XY-режиме. Необходимо отметить, что константой связи для системы вихрей является не $t$ (как для спиновых волн), а $y = \exp(-E_0/T)$ где $E_0$ – энергия вихря. Выбирая масштаб $\mu_1 \ll \sqrt{\eta}$, на котором осуществляется переход от уравнений (3.6) к (3.10) и используя уравнение сепаратрисы

$$8\pi y_1 = 1/t_1 - 4, \, t = t_{BKT}, \tag{3.11}$$

отделяющей низко-— и высокотемпературные фазы ($t_1 \equiv t_{\mu_1}$, $y_1 \equiv y_{\mu_1}$), можно воспроизвести результат (3.9) для температуры Березинского-Костерлица-Таулеса.

Уравнения (3.4) позволяют также определить температурную зависимость корреляционной длины выше температуры Березинского-Костерлица-Таулеса. В критической области выше $t_{BKT}$, т.е. при

$$\frac{1}{8\pi}(t_{BKT}^{-1} - t^{-1}) \ll 1, \qquad (3.12)$$

выражение для корреляционной длины имеет вид

$$\xi \simeq \frac{1}{\sqrt{\eta}} \exp\left(\frac{A}{2\sqrt{t_{BKT}^{-1} - t^{-1}}}\right), \qquad (3.13)$$

сходный с результатом для классической $XY$ модели ($A$ – некоторая константа). При условии, обратном (3.12), имеет место стандартное гейзенберговское поведение [46]

$$\xi = (C_\xi/\mu_0) t \exp(1/t). \qquad (3.14)$$

В присутствии межплоскостного обмена при достаточно низких температурах возникает магнитный порядок. Из-за топологических эффектов температура перехода при малом межплоскостном обмене стремится, однако, к $T_{BKT}$, а не к нулю. Описание РГ-преобразования при температурах вблизи критической области затруднительно вследствие сложной геометрии вихревых петель в трехмерном пространстве. Вместо прямого вычисления РГ траекторий можно использовать те же самые аргументы, что и в случае «легкая ось». Температура перехода определяется из требования, чтобы корреляционная длина модели без межплоскостного обмена ($\alpha = 0$) совпадала с характерным масштабом перехода от двумерной к трехмерной $XY$ модели. Тогда для критической температуры $t_c = T_C/(2\pi JS^2)$ (или $T_N/(2\pi\rho_s)$) находим при $\alpha \ll \eta$ [77]

$$t_c = \left\{\ln\frac{\mu_0}{\sqrt{\eta}} + 2\ln\frac{2}{t_{BKT}} + C - \frac{A^2}{\ln^2(\eta/\alpha)}\right\}^{-1}. \qquad (3.15)$$

Последний член в знаменателе определяет разницу между $t_c$ и $t_{BKT}$. Так как этот член может быть не слишком мал, разложение результата (3.15) по нему не производится.

Результат (3.15) качественно правилен вплоть до $\alpha$ порядка $\eta$ (в этом случае последний член в знаменателе приводит только к перенормировке константы $C$). В обратном пределе $\alpha \gg \eta$ поправки к результату РГ для квазидвумерных магнетиков вследствие анизотропии «легкая плоскость» определяются как [77]

$$t_c = \left[\ln\frac{\mu_0}{\sqrt{\alpha}} + 2\ln\frac{2}{t_c} + C' + O\left(\frac{\eta^{1/\psi}}{\alpha^{1/\psi}}\right)\right]^{-1}, \qquad (3.16)$$

где $\psi = \nu_3(2-\gamma_\eta)$ – критический индекс области, переходной между изотропным и анизотропным поведением, $\nu_3$ – соответствующий критический индекс трехмерной модели Гейзенберга, $\gamma_\eta$ – аномальная размерность параметра анизотропии трехмерной модели Гейзенберга. Результат $\varepsilon$–разложения в анизотропной модели $\phi^4$ в размерности $4-\varepsilon$ при $\varepsilon = 1$ есть $\psi \simeq 0.83$ [67]. Для антиферромагнетика согласно (2.90) постоянная $C' \simeq -0.066$. В отличие от (3.15), последний член в знаменателе (3.16) имеет степенную зависимость от параметра анизотропии. Это есть следствие того факта, что корреляционная длина в трехмерной модели Гейзенберга имеет степенную зависимость от температуры вблизи магнитного фазового перехода. По этой причине поправка в знаменателе (3.16) мала и для малой анизотропии ей можно пренебречь.

Обратимся теперь к экспериментальной ситуации. Наиболее экспериментально исследованная система с анизотропией типа «легкая плоскость» – соединение $K_2CuF_4$ является ферромагнетиком со спином $S = 1/2$, $T_{BKT} = 5.5\,\text{K}$, $T_C = 6.25\,\text{K}$ и параметрами $J = 20\,\text{K}$, $\eta = 0.04$, $\alpha = 6\cdot10^{-4}$ [4]. При подстановке этих значений в (3.9) и (3.15) можно определить $C \simeq -0.5$ и $A \simeq 3.5$. Эти значения констант могут быть проверены на других системах.

Другой пример квазидвумерного ФМ с анизотропией «легкая плоскость» – соединение $NiCl_2$ с $S = 1$. Согласно [4] его параметры – $J = 20 K$, $\eta = 8 \cdot 10^{-3}$ и $\alpha = 5 \cdot 10^{-5}$. Используя значения $A$ и $C$, определенные для $K_2CuF_4$, находим $T_{KT} = 17.4$ К и $T_C = 18.7$ К в согласии с экспериментальными данными (оба значения $T_{KT}$ и $T_C$ лежат в области $18 - 20$ К). В то же время вычисления с ведущей логарифмической точностью согласно (3.2) приводят к $T_{KT} = 35.3 K$, что вдвое больше экспериментального значения.

Соединение $BaNi_2(PO_4)_2$ согласно [4] является антиферромагнетиком с $S = 1$, $|J| = 22.0 K$ и анизотропией $\eta = 0.05, \alpha = 1 \cdot 10^{-4}$. Вычисление дает [86] $T_{KT} = 23.0 K$, что совпадает с экспериментальным значением и $T_N = 24.3 K$, снова в хорошем согласии с $T_N^{\exp} = 24.5 \pm 1 K$. Несмотря на то, что для этого соединения $T_{KT} \sim |J|S$, этот случай также должен рассматриваться как квантовый в соответствии с критерием квантового режима $(T/JS)^2 \ll 32$ (см. раздел 2.2).

# 4. Квазиодномерные изотропные антиферромагнетики

## 4.1. Модель и ее бозонизация

Хотя физическая ситуация для квазиодномерных магнетиков существенно отличается от квазидвумерного случая, при их теоретическом описании может быть также использована модель Гейзенберга (2.1). Ниже рассматривается простейший случай изотропных антиферромагнетиков ($\eta = \zeta = 0$) со спином $S = 1/2$ и малым межцепочечным обменом $|J'| \ll J$. При этом гамильтониан удобно записать в виде

$$H = J \sum_{n,i} \mathbf{S}_{n,i} \mathbf{S}_{n+1,i} + \frac{1}{2} J' \sum_{n,<ij>} \mathbf{S}_{n,i} \mathbf{S}_{n,j}, \qquad (4.1)$$

где $n$ нумерует узлы в цепочке, $i, j$ – индексы цепочек, $J > 0$ и $J'$ – внутри- – и межцепочечный обменные интегралы соответственно.

При исследовании элементарных возбуждений модели спиновые операторы в каждой цепочке могут быть представлены в терминах бозе-операторов $\varphi_i(x)$ (так называемая процедура "бозонизации"). В методе бозонизации спиновые операторы вначале представляются через ферми-операторы с помощью так называемого преобразования Йордана-Вигнера (см., напр., [78]). Указанное преобразование позволяет свести модель Гейзенберга к системе взаимодействующих фермионов. При этом поперечная часть спинового обмена приводит к гамильтониану свободных фермионов, а продольная часть отвечает их взаимодействию. Результирующие ферми операторы представляются через новые бозе-операторы $\varphi_i$ с помощью соотношений, позволяющих воспроизвести коммутационные соотношения исходных ферми-операторов. В результате находим выражения для исходных спиновых операторов через бозонные в виде [78]

$$\mathbf{S}_{n,i} = \mathbf{J}_i(x) + (-1)^n \mathbf{n}_i(x), \qquad (4.2)$$

где

$$J_i^z(x) = \frac{\beta}{2\pi} \partial_x \varphi_i(x),$$

$$J_i^\pm(x) = \frac{\lambda}{\pi} \exp[\pm i\beta \theta_i(x)] \cos \beta \varphi_i(x), \qquad (4.3)$$

- так называемые однородные компоненты спиновых операторов и

$$n_i^z(x) = \frac{\lambda}{\pi} \cos \beta \varphi_i(x),$$

$$n_i^\pm(x) = \frac{\lambda}{\pi} \exp[\pm i\beta \theta_i(x)], \qquad (4.4)$$

- соответствующие подрешеточные компоненты, $\lambda$ – постоянная масштаба обратной постоянной решетки, $\beta = \sqrt{2\pi}$, $\theta_i$ удовлетворяет соотношению $\partial_x \theta_i = -\Pi_i$, где $\Pi_i$ – импульс, канонически сопряженный с $\varphi_i$.

Гамильтониан (4.1), записанный в терминах бозе-операторов $\varphi_i(x)$ имеет вид

$$H = \frac{v}{2}\sum_i \int dx \left[ \Pi_i^2 + (\partial_x \varphi_i)^2 \right] + g_u \sum_i \int dx \cos 2\beta\varphi_i$$
$$- \frac{J'\lambda^2}{2\pi^2} \sum_{i,\delta_\perp} \int dx \left[ \cos(\beta\varphi_i)\cos(\beta\varphi_{i+\delta_\perp}) + \cos\beta(\theta_{i+\delta_\perp} - \theta_i) \right], \qquad (4.5)$$

где $v = \pi J/2$. Первая строка в (4.5) соответствует системе отдельных цепочек и имеет форму гамильтониана стандартной модели синус-Гордона. Первый член в (4.5) описывает свободную бозе-систему, а второй соответствует взаимодействию бозонов вдоль цепочек. Последнее возникает из-за рассеяния с процессом переброса («Umklapp» рассеяния) в системе фермионов, введенных преобразованием Йордана-Вигнера; это взаимодействие является маргинальным с РГ точки зрения и дает логарифмические поправки к термодинамическим величинам [36,79-83]. Численные оценки (см. [36,79]) приводят к значению $g_u/(2\pi) \simeq 0.25$. Вторая строчка в (4.5) описывает взаимодействие спинов между цепочками.

## 4.2. Приближение межцепочечного среднего поля для бозонизированного гамильтониана

Простейший способ рассмотрения межцепочечного обменного взаимодействия – так называемое межцепочечное приближение среднего поля [36]. Расцепляя член взаимодействия согласно

$$\cos(\beta\varphi_i)\cos(\beta\varphi_{i+\delta_\perp}) \to 2\langle\cos(\beta\varphi_{i+\delta_\perp})\rangle \cos(\beta\varphi_i), \qquad (4.6)$$

находим

$$H_{MF} = \frac{v}{2}\sum_i \int dx \left[ \Pi_i^2 + (\partial_x \varphi_i)^2 \right] + g_u \sum_i \int dx \cos 2\beta\varphi_i - \frac{\lambda}{\pi} h_{MF} \sum_i \int dx \cos(\beta\varphi_i) \quad (4.7)$$

где

$$h_{MF} = z_\perp J'\lambda \langle \cos(\beta\varphi_i) \rangle / \pi, \qquad (4.8)$$

$z_\perp$ – число ближайших соседей в поперечном к цепочке направлении ($z_\perp = 4$ для тетрагональной решетки). Приближение (4.6) дает возможность свести проблему многих цепочек к проблеме одной цепочки в эффективном подрешеточном магнитном поле. Вводя функцию

$$B(h;T) = \frac{\lambda}{\pi} \langle \cos(\beta \varphi_i) \rangle_h, \qquad (4.9)$$

вычисляемую в присутствии магнитного поля (последний член в (4.7)), получаем самосогласованное уравнение для подрешеточной намагниченности $\overline{S}$

$$\overline{S}_{MF} = B(z_\perp J' \overline{S}_{MF}; T). \qquad (4.10)$$

Несмотря на то, что гамильтониан $H_{MF}$ имеет одноцепочную форму, вычисление функции $B(h;T)$ (являющейся аналогом функции Бриллюэна в обычной теории среднего поля гейзенберговских магнетиков) при произвольных температурах — достаточно сложная задача. Согласно размерной оценке, $B(h;T) = h^{1/3} f(h^{2/3}/T)$ с некоторой функцией $f(x)$, $f(x) \sim x$ при $x \to 0$ и $f(\infty) =$ const. Для $g_u = 0$ (в этом случае имеем стандартную модель синус-Гордона или, что эквивалентно, массивную модель Тирринга) $B(h;T)$ была определена с помощью Бете-анзаца [84]. При $h \to 0$

$$B(h,T) = h\chi_0(T), \qquad (4.11)$$

где $\chi_0(T)$ — подрешеточная восприимчивость системы в отсутствии поля $h$ [36,83],

$$\chi_0(T) = \frac{\tilde{\chi}_0}{T} L\left(\frac{\Lambda J}{T}\right), \tilde{\chi}_0 = \frac{\Gamma^2(1/4)}{4\Gamma^2(3/4)} \simeq 2.1884, \qquad (4.12)$$

$$L(\Lambda J/T) = C\left[\ln\frac{\Lambda J}{T} + \frac{1}{2}\ln\ln\frac{\Lambda J}{T} + O(1)\right]^{1/2}. \qquad (4.13)$$

Константы $C$ и $\Lambda$ могут быть определены на основании численных расчетов [85]: $C \simeq 0.137$, $\Lambda \simeq 5.8$.

Результат (4.11) дает возможность вычислить значение $T_N$ в теории среднего поля, поскольку $h_{MF} \to 0$ при $T \to T_N$. Уравнение для температуры Нееля имеет вид [36]

$$T_N^{MF} = z_\perp J' \tilde{\chi}_0 L(\Lambda J/T_N^{MF}). \qquad (4.14)$$

Таким образом, согласно межцепочечной теории среднего поля $T_N \propto |J'|$; подрешеточная намагниченность основного состояния $\overline{S}_0 \propto \sqrt{|J'|/J}$ также зависит степенным образом от $J'$, что означает возникновение дальнего порядка при произвольно малых $|J'|$. Эти результаты противоречат стандартной теории спиновых волн, которая не делает различия между целыми и полуцелыми значениями спинов и предсказывает конечное критическое значение $J'_c \sim Je^{-\pi S}$ [30,53], так что при $|J'| < J'_c$ подрешеточная намагниченность $\overline{S}_0$ исчезает и

$$\overline{S}_0 \propto \ln|J'/J'_c|, \quad T_N \propto \overline{S}_0 \sqrt{|J'|} \qquad (4.15)$$

при $|J'| > J'_c$. Указанное противоречие было разрешено с помощью метода ренормгруппы [33-35], показавшего, что на масштабе обратной длины $\mu \gg J'_c/J$ стандартная спин-волновая теория действительно применима, причем перенормировочный фактор намагниченности $Z_\mu^{-1/2} \propto \ln \mu$. С другой стороны, для полуцелых спинов при $\mu \ll J'_c/J$ имеет место зависимость $Z_\mu^{-1/2} \propto \mu^{1/2}$ [33,34], означающая справедливость результатов теории межцепочечного среднего поля при $|J'| \ll J'_c$.

В то же время численные значения температуры Нееля в межцепочечной теории среднего поля оказываются сильно завышенными по сравнению с экспериментальными данными, поскольку эта теория не принимает во внимание эффекты корреляций между спинами, расположенными на разных цепочках. В частности, значение температуры Нееля (4.14) не чувствительно к пространственной размерности системы, хотя в двумерном случае должно быть $T_N = 0$; в трехмерном случае значения $T_N$ оказываются слишком высокими по сравнению с экспериментальными данными.

Корреляции между положениями спинов на разных цепочках выражаются в наличии коллективных возбуждений, вносящих вклад в термодинамические свойства. При этом ситуация в межцепочечной теории среднего поля аналогична недостаткам теории Стонера для зонных магнетиков, которая пренебрегает вкладом коллективных возбуждений, позже учтенных в теории Мории [24]. Как и в теории Мории, коллективные возбуждения в модели Гейзенберга могут быть рассмотрены в рамках приближения случайных фаз (ПСФ), в котором они определяются полюсами спиновых восприимчивостей [36,80]

$$\chi^{+-}(q_z,\omega) = \frac{\chi_0^{+-}(q_z,\omega)}{1 - J'(q_x,q_y)\chi_0^{+-}(q_z,\omega)/2}, \qquad (4.16)$$

$$\chi^{zz}(q_z,\omega) = \frac{\chi_0^{zz}(q_z,\omega)}{1 - J'(q_x,q_y)\chi_0^{zz}(q_z,\omega)}, \qquad (4.17)$$

где для тетрагональной решетки

$$J'(q_x,q_y) = 2J'(\cos q_x + \cos q_y), \qquad (4.18)$$

$\chi_0(q,\omega)$ – динамическая подрешеточная восприимчивость в модели (4.7). При $h \to 0$ восприимчивость $\chi_0(q,\omega)$ также определяется простыми аналитическими выражениями [82,83]:

$$\chi_0(q_z,\omega) = \frac{1}{T} L\left(\frac{\Lambda}{T}\right) \tilde{\chi}_0(q_z/T, \omega/T),$$

$$\tilde{\chi}_0(k,\nu) = \frac{1}{4} \frac{\Gamma(1/4 + ik_+)\Gamma(1/4 + ik_-)}{\Gamma(3/4 + ik_+)\Gamma(3/4 + ik_-)}, \; k_\pm = \frac{\nu \pm k}{4\pi}. \qquad (4.19)$$

При этом $\chi_0(0,0) = \chi_0(T)$.

Чтобы определить поправки к межцепочечной теории среднего поля, связанные с вкладом коллективных возбуждений, можно использовать $1/z_\perp$ – разложение ($z_\perp$ – число ближайших соседей в направлениях поперечных к цепочкам) [86]. Этот подход подобен $1/z$ разложению (или разложению по обратному радиусу взаимодействия), использовавшемуся много лет назад для улучшения стандартной теории среднего поля гейзенберговских магнетиков [87,88]; он позволяет определить температуру Нееля квазиодномерных систем с большей точностью, чем в межцепочечном приближении среднего поля.

## 4.3. Поправки первого порядка по $1/z_\perp$ к межцепочечному приближению среднего поля

Рассмотрим теорию возмущений по $J'/\max(h_{MF}, T) \sim 1/z_\perp$, являющуюся аналогом разложения по $J/\max(h_{MF}, T) \sim 1/z$ для трехмерных гейзенберговских магнетиков [88]. Для разложения подрешеточной намагниченности

$$\overline{S} = \lambda \langle \cos(\beta \varphi_i) \rangle_h / \pi \tag{4.20}$$

в ряд по $J'$ удобно использовать выражение для $\overline{S}$ в формализме континуального интеграла

$$\overline{S} = \frac{\lambda}{\pi} \frac{\int D\varphi \cos(\beta\varphi_i(0)) \exp(-L[\varphi])}{\int D\varphi \exp(-L[\varphi])}, \tag{4.21}$$

где $L[\varphi]$ – функция Лагранжа, соответствующая гамильтониану (4.5). В нулевом порядке по $J'$ (то есть при $J' = 0$) имеем

$$\overline{S}_0 = B(h; T), \tag{4.22}$$

где функция $B$ определена в (4.9). Разлагая (4.21) в ряд по $J'$, находим, что каждый член может быть представлен определенной диаграммой; диаграммная техника при этом совпадает с диаграммной техникой для спиновых операторов [87,88] (некоторые элементы диаграммной техники показаны на рис. 9).

Все диаграммы классифицируются согласно их порядку по $J'/\max(h_{MF}, T) \sim 1/z_\perp$. Диаграммы рис. 10, имеют нулевой порядок по $1/z_\perp$. Суммирование этих диаграмм приводит к сдвигу внешнего магнитного поля на величину среднего поля:

$$h \to \tilde{h} = h + h_{MF}, \quad h_{MF} = z_\perp J' \overline{S}. \tag{4.23}$$

(тот же самый результат мог быть получен исключением вклада среднего поля непосредственно из (4.21)). Диаграммы первого порядка по $1/z_\perp$ (см. рис. 11а) содержат одну линию взаимодействия ПСФ, являющегося суммой неприводимых диаграмм рис. 11б. В аналитическом виде это взаимодействие определяется как

$$V^{+-,zz}(\mathbf{q},\omega) = \frac{J'(q_x,q_y)}{1+\delta - J'(q_x,q_y)\chi_0^{+-,zz}(q_z,\omega)}, \quad (4.24)$$

где

$$\chi_0^{zz}(q_z,\omega) = \frac{\lambda^2}{\pi^2}\int d^2\mathrm{x}\,\langle\cos\beta\varphi_i(0)\cos\beta\varphi_i(\mathrm{x})\rangle_{0,ir}\exp(-iq_z x+i\omega_n\tau),$$

$$\chi_0^{+-}(q_z,\omega) = \frac{\lambda^2}{\pi^2}\int d^2\mathrm{x}\,\langle\exp\{i\beta[\theta_i(0)-\theta_i(\mathrm{x})]\}\rangle_0\exp(-iq_z x+i\omega_n\tau) \quad (4.25)$$

и

$$\langle AB\rangle_{ir} = \langle AB\rangle - \langle A\rangle\langle B\rangle \quad (4.26)$$

есть неприводимое среднее двух операторов. Аналогично теории Мории [24], в знаменатель (4.24) введена поправка к ПСФ $\delta = z_\perp J'\chi_0^{+-}(0,0)-1$, позволяющая удовлетворить теореме Голдстоуна, которая требует наличия полюса эффективного взаимодействия при q=0, ω=0 и $T \leq T_N$.

С учетом (4.24) для намагниченности подрешетки получается результат

$$\overline{S} = \frac{1}{T}h_{MF}\,\widetilde{\chi}_0 L\!\left(\frac{\Lambda}{T}\right)\!\left\{1+\frac{\pi^2}{2T\widetilde{\chi}_0}L\!\left(\frac{\Lambda}{T}\right)\!\int d^2\mathrm{r}V(\mathrm{r})\!\left[\frac{1}{8}F(\mathrm{r})+\frac{1}{2}G(\mathrm{r})\right]\right\}, \quad (4.27)$$

где

$$V(\mathrm{r}) = \int_{-\infty}^{\infty}\frac{dq_z}{2\pi}\sum_n\sum_{q_x,q_y}\frac{\cos q_x+\cos q_y}{2\widetilde{\chi}_0-(\cos q_x+\cos q_y)\widetilde{\chi}_0(q_z,2\pi in)}\exp(iq_z r-2\pi in\tau) \quad (4.28)$$

$$\widetilde{\chi}_0 = \frac{\pi}{2}\int d^2\mathrm{z}\,\frac{1}{|\widetilde{\varsigma}(\mathrm{z})|} \simeq 2.1184 \quad (4.29)$$

и функции $F(\mathrm{r})$, $G(\mathrm{r})$ определены в [86], $\tilde{\varsigma}(\mathrm{x}) = \sinh(\pi \mathrm{x})$. С использованием связи между средним полем и подрешеточной намагниченностью (4.23) после собирания всех поправок в знаменатель результат для температуры Нееля в первом порядке по $1/z_\perp$ принимает вид

$$T_N = kJ'z_\perp\tilde{\chi}_0 L(\Lambda/T_N). \quad (4.30)$$

Результат (4.30) отличается от результата тории среднего поля (4.14) множителем $k$, зависящем от структуры решетки в направлении,

перпендикулярном к цепочкам. Численный расчет для тетрагональной решетки приводит к значению $k \simeq 0.70$. Таким образом, уменьшение $T_N$ благодаря флуктуационным эффектам составляет 25% его средне-полевого значения, что находится в хорошем согласии с результатами численного анализа [89]. В двумерном случае имеем $k=0$, так что $T_N = 0$.

Поправки к подрешеточной намагниченности основного состояния могут быть вычислены аналогичным образом [86]. Для динамической восприимчивости одной цепочки при $T = 0$ имеем [36,80]

$$\chi_0^{+-} = \frac{1}{4|J'|} \frac{\Delta^2}{\omega^2 + v^2 q^2 + \Delta^2} \qquad (4.31)$$

$$\chi_0^{zz} = \frac{Z'/Z}{4|J'|} \frac{\Delta^2}{\omega^2 + v^2 q^2 + 3\Delta^2} \qquad (4.32)$$

где $\Delta \simeq 6.175|J'|$ – щель в спектре спиновых возбуждений, $Z, Z'$ – спектральный вес продольных и поперечных одномагнонных возбуждений ($Z'/Z \simeq 0.49$), $\overline{S}_0 \simeq 1.017|J'|$ - подрешеточная намагниченность в основном состоянии и $h_{MF} = z_\perp J' \overline{S}_0$. Используя вновь ПСФ для потенциала взаимодействия возбуждений на разных цепочках (4.27), находим

$$\overline{S} = \overline{S}_{MF} - \frac{\Delta}{4\pi} \frac{\partial \Delta}{\partial h_{MF}} I,$$

$$I = \sum_q \left[ (1-\Gamma'_q/2) \ln \frac{1}{1-\Gamma'_q} + (3 - Z'\Gamma'_q/2Z) \ln \frac{1}{1 - Z'\Gamma'_q/(3Z)} \right], \qquad (4.33)$$

где $\Gamma'_q = \cos q$ для двумерной и $\Gamma'_q = (\cos q_x + \cos q_y)/2$ для тетрагональной решетки. Численное интегрирование приводит к результату [86]

$$\overline{S}_0 = (0.677 - I) h_{MF}^{1/3}. \qquad (4.34)$$

Последний член в скобках в (4.34) представляет собой $1/z_\perp$-поправку к намагниченности основного состояния,

$$I = \begin{cases} 0.011 & (3D) \\ 0.060 & (2D) \end{cases}. \qquad (4.35)$$

Из (4.34) следует, что намагниченность основного состояния уменьшается почти на 10% по сравнению с ее значением в теории среднего поля для двумерной и только на 2% для трехмерной решеток. Таким образом, флуктуационные поправки для подрешеточной намагниченности основного состояния гораздо менее важны, чем для температуры Нееля, и в трехмерном случае ими можно пренебречь.

### 4.4. Сравнение с экспериментальными данными

Рассмотрим применение приведенных теоретических результатов к описанию экспериментальных данных для магнитных квазиодномерных систем. Наиболее изученным квазиодномерным соединением является $KCuF_3$, имеющее спин $S = 1/2$. Эксперименты нейтронного рассеяния [10] приводят к параметру магнитного обмена вдоль цепочек для этого соединения $J = 406$ K и намагниченности основного состояния $\overline{S}_0/S = 0.25$. Как обсуждается в [36], это значение $\overline{S}_0$ соответствует $J'/J = 0.047$, так, что $J' = 19.1$ K. Межцепочечное приближение среднего поля (4.14) приводит к значению $T_N = 47$ K при этих параметрах, что несколько выше экспериментального результата $T_N = 39$ K [10]. В то же время, результат $1/z_\perp$-разложения (4.30) $T_N = 37.7$ K находится гораздо ближе к экспериментальному значению. Таким образом, рассматриваемый подход слегка переоценивает флуктуационные эффекты, но значительно улучшает межцепочечное приближение среднего поля. Вклад двойного логарифмического члена в (4.13) составляет приблизительно 5% и улучшает согласие с экспериментальными данными.

Другое соединение с $S = 1/2$, широко обсуждаемое в литературе, – $Sr_2CuO_3$ – имеет следующие параметры [11,12]: $J = 2600$ K, $T_N = 5$ K. Надежные экспериментальные данные для $J'$ отсутствуют, но, используя (4.30) и экспериментальное значение $T_N$, находим $J' = 1.85$ K. Тогда из (4.34) следует $\overline{S}_0/S = 0.042$, что находится в согласии с экспериментальными данными ($\bar{S}_0/S \lesssim 0.05$).

Для $Ca_2CuO_3$ экспериментальные параметры имеют следующие значения [11,12]: $S = 1/2$, $J = 2600\,K$ и $T_N = 11\,K$. Из них находим $J' = 4.3$ K и $\bar{S}_0/S = 0.062$, что снова находится в хорошем согласии с экспериментальными данными [12], которые дают $\bar{S}_0(Ca_2CuO_3)/\bar{S}_0(Sr_2CuO_3) = 1.5 \pm 0.1$. Таким образом, результат (4.30) достаточен для количественного описания реальных квазиодномерных магнитных систем.

## 5. Заключение

Квазиодномерные и слоистые магнетики представляют собой пример систем с сильными флуктуациями и нетривиальным поведением термодинамических и магнитных свойств. Исследование этих систем – весьма нетривиальная проблема с точки зрения теоретической физики. Обычная спин-волновая теория (и даже ее усовершенствованный самосогласованный вариант – ССВТ), хотя и приводит к правильному результату для температуры перехода $T_М$ в ведущем логарифмическом приближении, оказывается количественно применимой лишь при температурах $T \ll T_М$. В области более высоких температур необходим учет динамического взаимодействия спиновых волн, выходящий за рамки низшего (борновского) приближения, а также существенно не спин-волновых возбуждений.

Проблема описания термодинамических свойств квазиодномерных и слоистых магнетиков получила существенное развитие в рамках теоретико-полевых методов, примененных к широко распространенной модели магнетизма этих систем – модели Гейзенберга. Использование этих подходов позволяет получить простые аналитические результаты для температурной зависимости намагниченности и величины $T_М$, которые могут быть использованы при практической обработке экспериментальных данных. В квазидвумерных магнетиках в широкой температурной области ниже $T_M$ спин-волновая картина спектра возбуждений является адекватной и взаимодействие спиновых волн приводит к появлению поправочных слагаемых в выражениях для намагниченности и обратной температуры Нееля $1/T_M$, значительно улучшающих согласие с экспериментальными данными. Узкая критическая область вблизи $T_M$ может быть описана с учетом неспинволновых возбуждений, в т.ч. в рамках $1/N$ разложения. В квазиодномерных магнетиках переход к бозевским (не спин-волновым) возбуждениям позволяет построить систематическое разложение по обратному координационному числу решетки в направлениях, перпендикулярных к цепочкам.

Таким образом, с теоретической точки зрения в последнее время достигнуто хорошее понимание физической картины спектра и свойств низкоразмерных магнетиков в широком интервале температур. Оно дает основу для количественного описания свойств реальных систем, и мы ставили одной из своих задач привлечь внимание экспериментаторов к этому факту. В то же время при детальном анализе магнетизма конкретных соединений необходим учет дипольного взаимодействия, релятивистских взаимодействий типа Дзялошинского-Мории и т.д. Несмотря на то, что уже имеются первые попытки описания систем с такими взаимодействиями в рамках самосогласованного спин-волнового и теоретико-полевого подходов [90, 91], они ждут своего дальнейшего развития. С другой стороны, широко исследуемые в последнее время комплексные соединения со сложной кристаллической структурой, а также системы типа пленок и мультислоев, рассматривавшиеся ранее в рамках спин-волновой теории [92], требуют более конкретного изучения в рамках описанных подходов.

Близкие проблемы возникают при описании систем, имеющих фрустрированные магнитные структуры – на двумерной квадратной решетке с учетом обменных взаимодействий между следующими за ближайшими соседями [93-97], двумерной треугольной решетке [98-103], решетках Кагоме, пирохлора, [104-105] и т.д. Наличие спиновых фрустраций в таких системах приводит, как и в низкоразмерных соединениях, к подавлению дальнего магнитного порядка (при сохранении ближнего) и, следовательно, к очень нетривиальным термодинамическим свойствам. Фрустрированные системы также рассматривались в рамках спин-волновых теорий [106-111].

Еще одна проблема, актуальная, например, в связи с высокотемпературной сверхпроводимостью и не затронутая в обзоре, – взаимодействие носителей тока с магнитными моментами. Специфика низкоразмерных систем (сильный ближний магнитный порядок) приводит к соответствующим особенностям электронного спектра [112,113]. Сильное электрон-электронное взаимодействие в этих условиях является дополнительным фактором, приводящим к формированию некогерентных электронных состояний и возможности перехода металл-изолятор. В связи с этим, сейчас ведется интенсивное теоретическое и экспериментальное исследование проводящих низкоразмерных систем, находящихся вблизи такого перехода [114-116]. Оно требует развитие существенно новых подходов, в которых, однако, могут быть использованы теоретические методы описания подсистемы локализованных моментов, изложенные в настоящем обзоре.

# Список литературы

1. Birgeneau R J, Guggenheim H J, Shirane G Phys. Rev. B **1** 2211 (1970).

2. de Jongh L J and Miedema A R, *Experiments on Simple Magnetic Model Systems* (London : Taylor and Francis, 1974).

3. Birgeneau R J Guggenheim H J, Shirane G Phys. Rev. B **8** 304 (1973).

4. *Magnetic Properties of Layered Transition Metal Compounds* (Ed. de Jongh L J) (Dordrecht: Cluwer, 1989).

5. Lahti P M *Magnetic properties of organic materials* (N. Y.: Marcel Dekker, 1999).

6. Blundell S J, Pratt F L J. Phys.: Cond. Matt. **16** R771 (2004).

7. Allenspach A J. Magn. Magn. Mater. **129** 160 (1994).

8. Elmers H.-J. Int. J. Mod. Phys. B **9** 3115 (1995).

9. Birgeneau R J, Gabbe D R, Jenssen H P, Kastner M A, Picone P J, Thurston T R, Shirane G, Endoh Y, Sato M, Yamada K, Hidaka Y, Oda M, Enomoto Y, Suzuki M, Murakami T Phys. Rev. B **38**, 6614 (1988).

10. Satija S K, Axe J D, Shirane G, Yoshizawa H, Hirakawa K Phys. Rev. B 21, 2001 (1980); Tennant D A Phys. Rev. B 52 13381 (1995).

11. Keren A, Le L P, Luke G M, Sternlieb B J, Wu W D, Uemura Y J, Tajima S, Uchida S Phys. Rev. B **48** 12926 (1993); Ami T, K. Crawford M, L. Harlow R, R. Wang Z, C. Johnston D, Huang Q, W. Erwin R Phys. Rev. B **51** 5994 (1995).

12. Kojima K M, Fudamoto Y, Larkin M, Luke G M, Merrin J, Nachumi B, Uemura Y J, Motoyama N, Eisaki H, Uchida S, Yamada K, Endoh Y, Hosoya S, Sternlieb B J, Shirane G Phys. Rev. Lett **78** 1787 (1997).

13. Kenzelmann M, Cowley R A, Buyers W J L, Coldea R, Gardner J S, Enderle M, McMorrow D F, Bennington S M Phys. Rev. Lett **87** 017201 (2001); Kenzelmann M, Cowley R A, Buyers W J L, Tun Z, Coldea R, Enderle M Phys. Rev. B **66** 024407 (2002).

14. Kadowaki H, Hirakawa K, Ubukoshi K J. Phys. Soc. Jpn. **52** 1799 (1983); Itoh S, Kakurai K, Arai M, Endoh Y J. Phys.: Cond. Matt. **5** 6767 (1993).

15. Dagotto E, Rice T M Science **271** 618 (1996).

16. Anderson P W Phys. Rev. **86** 694 (1952).

17. Dyson F Phys. Rev. **102** 1217 (1956); **102** 230 (1956).

18. Малеев С В ЖЭТФ **33** 1010 (1957).

19. Harris A B, and D. Kumar, B. I. Halperin and P. C. Hohenberg Phys. Rev. B **3** 961 (1971)

20. Tyc S, Halperin B I Phys. Rev. B **42** 2096 (1990).

21. Каганов М И, Чубуков А УФН **153** 537 (1987); Косевич Ю А, Чубуков А ЖЭТФ 91 1105 (1990).

22. Барьяхтар В Г, Криворучко В Н, Яблонский Д А ЖЭТФ **85** 602 (1983).

23. Барьяхтар В Г, Криворучко В Н, Яблонский Д А *Функции Грина в теории магнетизма* (Киев: Наукова Думка, 1984).

24. Мория Т *Спиновые флуктуации в магнетиках с коллективизированными электронами* (М: Мир, 1988).

25. Паташинский А З, Покровский В Л *Флуктуационная теория фазовых переходов* (М.: Наука, 1982).

26. Д. Ландау Л, Лифшиц Е М *Электродинамика сплошных сред* (М.: Наука, 1982).

27. Chubukov A V, Sachdev S, Ye J Phys. Rev. B **49** 11919 (1994).

28. Irkhin V Yu, Katanin A A Phys. Rev. B **55** 12318 (1997).

29. Haldane F D M Phys. Lett. A **93** 464 (1983); Phys. Rev. Lett. **50** 1153 (1983).

30. Affleck I J. Phys.: Cond. Matt. **1** 3047 (1989).

31. Shelton D G, Nersesyan A A, Tsvelik A M Phys. Rev. B **53** 8521 (1996).

32. Hori H, Yamamoto S J. Phys. Soc. Jpn. 73 **3** (2004).

33. Affleck I, Gelfand M P, Singh R R P J. Phys. A **27** 7313 (1994).

34. Affleck I, Halperin B I J. Phys. A **29** 2627 (1996).

35. Wang Z Phys. Rev. Lett. **78** 126 (1997).

36. Schulz H Phys. Rev. Lett. **77** 2790 (1996).

37. Маттис Д *Теория магнетизма* (М.: Мир, 1967).

38. Arovas D P, Auerbach A Phys. Rev. B **38** 316 (1988).

39. Yoshioka D J J. Phys. Soc. Jpn. **58** 3733 (1989).

40. Holstein T, Primakoff H Phys. Rev. **58** 1098 (1940).

41. Loly P D J. Phys. C **1** 1365 (1971).

42. Bloch M Phys. Rev. Lett **9** 286 (1962).

43. Rastelli E, Tassi A, Reatto L J. Phys. C **7** 1735 (1974).

44. Sarker S Phys. Rev. B **40** 5028 (1989).

45. Takahashi M Phys. Rev. B **40** 2494 (1989).

46. Chakravarty S, Halperin B I, Nelson D R Phys. Rev. B **39** 2344 (1989).

47. Kopietz P, Chakravarty S Phys. Rev. B **40** 4858 (1989).

48. Irkhin V Yu, Katanin A A, Katsnelson M I, Phys. Lett. A **157** 295 (1991); ФММ **79** №1 65 (1995).

49. Kopietz P Phys. Rev. Lett. **68** 3480 (1992).

50. Liu Bang-Gui, J. Phys. Cond. Matt **4** 8339 (1992).

51. Irkhin V Yu, Katanin A A, Katsnelson M I Phys. Rev. B **60** 1082 (1999).

52. Барабанов А Ф, Старых О А Письма в ЖЭТФ **51** 271 (1991).

53. Irkhin V Yu, Katanin A A, Katsnelson M I J. Phys. Cond. Matt. **4** 5227 (1992).

54. Xu J H, Ting C S Phys. Rev. B **42** 6861 (1990).

55. Oguchi T, Kitatani H J. Phys. Soc. Japan **59** 3322 (1990).

56. Nishimori H, Saika Y J. Phys. Soc. Japan **59** 4454 (1990).

57. Тябликов С В *Методы квантовой теории магнетизма* (М.: Наука, 1975).

58. Irkhin V Yu, Katanin A A, Katsnelson M I Phys. Rev. B **54** 11953 (1996).

59. Chubukov A V, Starykh O A Phys. Rev. B **52** 440 (1995).

60. Polyakov A M Phys. Lett. B **79** (1975).

61. Brezin E, Zinn-Justin J Phys. Rev. B **14** 3110 (1976).

62. Nelson D R, Pelkovitz R A Phys. Rev. B **16** 2191 (1977).

63. Irkhin V Yu, Katanin A A  Phys. Rev. B **57** 379 (1998).

64. Нагаев Э Л *Магнетики со сложными обменными взаимодействиями* (М.: Наука, 1988).

65. Klauder J R  Phys. Rev. D **19** 2349 (1979).

66. Auerbach A *Interacting Electrons and Quantum Magnetism* (New York: Springer-Verlag, 1994).

67. Amit D *Field Theory, the Renormalization Group, and Critical Phenomena* (Singapore: World Scientific, 1984).

68. Ма Ш *Современная теория критических явлений* (М.: Мир, 1980).

69. Keimer B, Aharony A, Auerbach A, Birgeneau R J, Cassanho A, Endoh Y, Erwin R W, Kastner M A, Shirane G  Phys. Rev. B **45** 7430 (1992).

70. Aeppli G, Hayden S M, Mook H A, Fisk Z, Cheong S W, Rytz D, Remeika J P, Espinosa G P, Cooper A S  Phys. Rev. Lett. **62** 2052 (1989); Lyons K B, Fleury P A, Remeika J P, Cooper A S, Negran T J Phys. Rev. B **37** 2353 (1988).

71. Katanin A A, Kampf A P Phys. Rev. B **66** 100403 (2002).

72. Levanjuk  A, Garcia N,  J. Phys.: Cond. Matt. **4** 10277 (1992); Serena P A, Garcia N, Levanjuk A Phys. Rev. B **47** 5027 (1993).

73. Березинский В Л ЖЭТФ **59** 907 (1970).

74. Kosterlitz  J M, Thouless D J J. Phys. C **6** 1181 (1973); Kosterlitz J M J. Phys. C **7** 1046 (1974).

75. Jose J V, Kadanoff L P, Kirpatrick S, Nelson D R Phys. Rev. B **16** 1217 (1977).

76. Hikami S, Tsuneto T Progr. Theor. Phys. **63** 387 (1980).

77. Irkhin V Yu, Katanin A A  Phys. Rev. B **60** 2990 (1999).

78. Tsvelik A M *Quantum Field Theory in Condensed Matter Physics* (Cambridge: Cambridge University Press, 1995).

79. Affleck I, Gepner D, Schulz H J, Ziman T  J. Phys. A **22** 511 (1989).

80. Essler F H L, Tsvelik A M, Delfino G  Phys. Rev. B **56** 11001 (1997).

81. Barzykin V, Affleck I  J. Phys. A **32** 867 (1999).

82. Schulz H J  Phys. Rev. B **34** 6372 (1986).

83. Barzykin V препринт cond-mat/9904250.

84. Chung S G, Chang Y C  J. Phys. A **20** 2875 (1987).

85. Starykh O A, Sandvik A W, Singh R R P  Phys. Rev. B **55** 14953 (1997).

86. Irkhin V Yu, Katanin A A  Phys. Rev. B **61** 6757 (2000).

87. Вакс В Г, Ларкин А И, Пикин С А ЖЭТФ **53** 281 (1967).

88. Изюмов Ю А, Кассан-Оглы Ф А, Скрябин Ю Н *Полевые методы в теории ферромагнетизма* (М.: Наука, 1974); Изюмов Ю А, Скрябин Ю Н *Статистическая механика магнито-упорядоченных систем* (М.: Наука, 1987).

89. Yasuda C, Todo S, Hukushima K, Alet F, Keller M, Troyer M, Takayama H Phys. Rev. Lett. **94** 217201 (2005).

90. Grechnev A, Irkhin V Yu, Katsnelson M I, Eriksson O Phys. Rev. B **71** 024427 (2005).

91. Benfatto L, Silva Neto M, Juricic V, Morais Smith C, Physica B **378** 449 (2006); Silva Neto M, Benfatto L, Juricic V, Morais Smith C, Phys.Rev. B **73** 045132 (2006).

92. Irkhin V Yu, Katanin A A, Katsnelson M I JMMM **140-144** 1695 (1995).

93. Ballon R, Lacroix C, Nunez M D Phys. Rev. Lett. **66** 1910 (1991).

94. Chattopadhyay T, Bruchel T, Burlet P  Phys. Rev. B **44** 7394 (1991).

95. Chubukov A V, J. Phys.: Cond. Matt. **2** 4455 (1990).

96. Rastelli E, Tassi A Phys. Rev. B **44** 7135 (1991).

97. Ferrer J Phys. Rev. B **47** 8769 (1993).

98. Тext P C УФН **159** 261 (1989).

99. Hirota K, Nakazawa Y, Ishikawa M, Tech. Rep. ISSP A 2286 (1990).

100. Yoshizawa H, Mori H, Hirota K, Ishikawa M Tech. Rep. ISSP A 2289 (1990).

101. Chubukov A V, Jolicur Th Phys. Rev. B **46** 11137 (1992).

102. Korshunov S E Phys. Rev. B **43** 6165 (1993).

103. Yang K, Warman L K, Girvin S M, Phys. Rev. Lett. **70** 2641 (1993).

104. Reimers J N, Gredau J E, Kremar R K, Gmedin E, Subramanian M A Phys. Rev. B **43** 3387 (1991); Reimers J N, Greedan J E, Stager C V, Bjargvinnsen M, Subramanian M A Phys. Rev. B **43** 5692 (1991).

105. Gaulin B D, Reimers J N, Matson T E, Greeden J E, Tun Z Phys. Rev. Lett. **63** 3244 (1992)

106. Xu J H, Ting C S, Phys. Rev. B **42** 6861 (1990).

107. Nishimori H, Saika Y J. Phys. Soc. Jpn. **53** 4454 (1990)

108. Jolicur Th, Guillom J. Phys. Rev. B **40** 2727 (1989).

109. Yoshioka D, Miyazaki Y, J. Phys. Soc. Jpn. **60**, 614 (1991).

110. Hizi U, Sharma Prashant, Henley C L Phys. Rev. Lett. **95** 167203 (2005); Hizi U, Henley C L Phys. Rev. B **73** 054403 (2006).

111. Del Maestro A G, Gingras M J P J. Phys.: Cond. Mat. **16** 3339 (2004).

112. Irkhin V Yu, Katsnelson M I J. Phys.: Cond. Mat. **3** 6439 (1991); Phys. Rev. B **62** 5647 (2000).

113. Irkhin V Yu, Katsnelson M I Phys. Rev. B**53**, 14008 (1996); Europ. Phys. Journal B **19** 401 (2001).

114. Maier T A, Pruschke T, Jarell M Phys. Rev. B **66** 075102 (2002).

115. Senechal D, Tremblay A M S Phys. Rev. Lett. **92** 126401 (2004).

116. Civelli M, Capone M, Kancharla S S, Parcollet O, Kotliar G Phys. Rev. Lett. **95** 106402 (2005).

# Magnetic order and spin fluctuations in low-dimensional systems

A. A. Katanin[a,b] and V. Yu. Irkhin[a]

a) Institute of Metal Physics, Urals Division RAS,
Kovalevskaya st. 18, 620041, Ekaterinburg, Russia
Tel. +7(343)3783778
Fax +7(343)3745244
Valentin.Irkhin@imp.uran.ru

b) Max-Planck Institut für Festkörperforschung,
Heisenberg str. 1, 70569 Stuttgart, Germany
Tel. +49(0711)6891536
Fax +49(0711)6891702
A.Katanin@fkf.mpg.de

We analyze the modern theoretical and experimental situation in the physics of low-dimensional insulating systems which have low values of magnetic transition temperature and demonstrate pronounced short-range magnetic order above this temperature. The insufficiency of the standard and self-consistent spin-wave theories for the quantitative description of experimental data on these systems is shown. The field-theoretical approaches which enable one to take into account the contribution of spin-fluctuation excitations to thermodynamic properties of ferro- and antiferromagnets, not considered in spin-wave theories are discussed.

Подписи к рисункам:

Рис. 1. Температурная зависимость намагниченности квазидвумерных ферромагнетиков при разных значениях отношения обменных интегралов между плоскостями и в плоскости $J'/J$ ($S = 1/2$).

Рис. 2. Зависимость параметра ближнего порядка γ от температуры при тех же значениях параметров, что и на Рис. 1.

Рис. 3. Зависимость щели в спектре бозонов от температуры при тех же значениях параметров, что и на Рис. 1.

Рис. 4. Теоретические температурные зависимости относительной намагниченности подрешетки $\bar{\sigma}_r$ в различных приближениях: спин-волновых теориях, РГ подходе (2.105) и $1/N$ разложении в $O(N)$ модели (уравнения (2.110) и (2.111)) и экспериментальные точки для La$_2$CuO$_4$ [3]. Кривая РГ приведена вплоть до температуры, где производная $\partial \bar{\sigma}_r/\partial T$ расходится. Кривая, обозначенная как $1/N'$, ближе к экспериментальным данным в переходной температурной области благодаря включению анизотропии, определенной из условия равенства $T_M$ его экспериментальному значению (см. обсуждение в тексте).

Рис. 5. Относительная намагниченность подрешетки $\bar{\sigma}_r(T)$ для K$_2$NiF$_4$ (точки) по сравнению со стандартной спин-волновой теорией (пунктир), ССВТ (штрих-пунктир), РГ подходом и результатом решения уравнения (2.110) в промежуточной температурной области (сплошная линия). Короткий пунктир показывает экстраполяцию результата $1/N$ разложения на изинговскую критическую область. Граница между областью с флуктуациями двумерного типа и поведения переходного к критическому отмечена стрелкой.

Рис. 6. Экспериментальная зависимость $\bar{\sigma}_r(T)$ для K$_2$MnF$_4$ (точки) по сравнению с результатами ССВТ (пунктирная линия), квантовым РГ анализом (две точки-пунктир), классическим РГ анализом (штрих-пунктир) и решением (2.110) (сплошная линия).

Рис. 7. Результаты ренормгруппового подхода (сплошная линия) и ССВТ (пунктирная линия) для относительной намагниченности $\bar{\sigma}$ классического анизотропного двумерного магнетика ($\zeta = 0$, $\eta = 0.001$) в сравнении с результатамм вычисления методом Монте-Карло [72]. РГ и ССВТ кривые показаны до температуры, где $\partial\bar{\sigma}/\partial T = \infty$.

Рис. 8. Схематическая картина РГ траекторий в слоистых магнетиках. Левая сторона: преобразование от двумерной модели Гейзенберга с анизотропией «легкая ось» (Н+ЛО) к двумерной модели Изинга. Правая сторона: преобразование от двумерной модели Гейзенберга с анизотропией «легкая плоскость» (Н+ЛП) к двумерной XY модели. Точки перегиба $c_1$, $c_2$ отмечают переходные области. Пунктирные линии – для соответствующих квазидвумерных моделей.

Рис. 9. Некоторые элементы диаграммной техники для спиновых операторов (см. детальное описание в [88]).

Рис. 10. Диаграммы для подрешеточной намагниченности в нулевом порядке по $1/z_\perp$ (приближение среднего поля).

Рис. 11. (а) Диаграммы первого порядка по $1/z_\perp$ для подрешеточной намагниченности (б) уравнения для ПСФ линий взаимодействия.